\RequirePackage{etex}
\documentclass[doublespace,onecolumn]{SSA-SRL-Ian} 

\usepackage{listings}
\usepackage{xcolor}  
\usepackage{caption}
\lstdefinestyle{pythonStyle}{
    language=Python,
    backgroundcolor=\color{white},
    basicstyle=\ttfamily\small,
    breaklines=true,
    captionpos=b,
    numbers=left,
    numberstyle=\tiny\color{gray},
    keywordstyle=\color{blue},
    commentstyle=\color{green},
    stringstyle=\color{red},
    showstringspaces=false
}

\citethisauthor{Denolle M., Tape C., Bozda\u g E., Wang Y., Waldhauser F., Gabriel A. A., Braunmiller J., Chow B., Ding L., Feng K.F., Ghosh A., Groebner N., Gupta A., Krauss Z., McPherson A., Nagaso M., Niu Z., Ni Y., \" Orsvuran R., Pavlis G., Rodriguez-Cardozo F., Sawi T., Schliwa N., Schneller D., Shi Q., Thurin J., Wang C., Wang K., Wong J. W. C., Wolf S., and Yuan C.}
\doi{00.0000/000000000}
\recdate{23 September 2024}

\begin{document}

\title{Training the Next Generation of Seismologists: Delivering Research-Grade Software Education for Cloud and HPC Computing through Diverse Training Modalities}

\author[*1\orc{0000-0002-1610-2250}]{Marine A. Denolle}
\author[2\orc{0000-0003-2804-7137}]{Carl Tape}
\author[3a,3b\orc{0000-0002-4269-3533}]{Ebru Bozda\u g}
\author[4\orc{0000-0001-8505-0223}]{Yinzhi Wang}
\author[5\orc{0000-0002-1286-9737}]{Felix Waldhauser}
\author[6,7\orc{0000-0003-0112-8412}]{Alice-Agnes Gabriel}

\author[11\orc{0000-0002-8541-6560}]{Jochen Braunmiller}
\author[2\orc{0000-0002-3901-4755}]{Bryant Chow}
\author[12\orc{0000-0002-3047-9556}]{Liang Ding}
\author[1\orc{0009-0002-0765-4499}]{Kuan-Fu Feng}
\author[3b\orc{0000-0001-9627-1849}]{Ayon Ghosh}
\author[11\orc{0000-0002-6535-1739}]{Nathan Groebner}
\author[2\orc{0000-0002-2776-6431}]{Aakash Gupta}
\author[1\orc{0000-0002-5933-6823}]{Zoe Krauss}
\author[2\orc{0000-0002-4382-8786}]{Amanda M. McPherson}
\author[3a\orc{0000-0002-3566-6174}]{Masaru Nagaso}
\author[7\orc{0000-0003-4213-3322}]{Zihua Niu}
\author[1\orc{0000-0001-5181-9700}]{Yiyu Ni}
\author[3a\orc{0000-0002-5098-7515}]{R\i dvan \" Orsvuran}
\author[13\orc{0000-0003-2128-9897}]{Gary Pavlis}
\author[11\orc{0000-0002-8456-8001}]{Felix Rodriguez-Cardozo}
\author[5,8\orc{0000-0001-5938-6743}]{Theresa Sawi}
\author[5\orc{}]{David Schaff}
\author[7\orc{0009-0004-7825-9458}]{Nico Schliwa}
\author[9\orc{0000-0002-2145-6133}]{David Schneller}
\author[1\orc{0000-0002-4211-9187}]{Qibin Shi}
\author[2\orc{0000-0001-9982-3768}]{Julien Thurin}
\author[4\orc{0009-0008-4031-1782}]{Chenxiao Wang}
\author[5\orc{0000-0001-7838-6984}]{Kaiwen Wang}
\author[6\orc{0000-0001-5523-8145}]{Jeremy Wing Ching Wong}
\author[9]{Sebastian Wolf}
\author[10\orc{0000-0002-1877-5539}]{Congcong Yuan}

\affil[1]{Department of Earth and Space Sciences, University of Washington, Johnson Hall 070, Box 351310, 1707 NE Grant Lane, Seattle, WA 98105, USA}{\auorc[https://orcid.org/]{0000-0002-1610-2250}{(MD)} \auorc[https://orcid.org/]{0000-0001-5181-9700}{(YN)}
\auorc[https://orcid.org/]{0000-0002-5933-6823}{(ZK)}
\auorc[https://orcid.org/]{0000-0002-1115-2427}{(KF)}
\auorc[https://orcid.org/]{0000-0002-4211-9187}{(QS)}}

\affil[3a]{Department of Applied Mathematics and Statistics, Colorado School of Mines, 1301 19th Street
Golden, CO 80401, USA}{\auorc[https://orcid.org/]{0000-0002-4269-3533}{(EB)} \auorc[https://orcid.org/]{0000-0002-3566-6174}{(MN)} \auorc[https://orcid.org/]{0000-0002-5098-7515}{(RO)}}
\affil[3b]{Department of Geophysics, Colorado School of Mines, 924 16th Street, Golden, CO 80401, USA}{\auorc[https://orcid.org/]{0000-0002-4269-3533}{(EB)}\auorc[https://orcid.org/]{0000-0001-9627-1849}{(AG)} }

\affil[4]{Texas Advanced Computing Center, The University of Texas at Austin, 10100 Burnet Rd, Austin, TX 78758, USA}{\auorc[https://orcid.org/]{0000-0001-8505-0223}{(YW)} \auorc[https://orcid.org/]{0009-0008-4031-1782}{(CW)}}

\affil[6]{Institute of Geophysics and Planetary Physics, Scripps Institution of Oceanography, University of California at San Diego, 9500 Gilman Drive La Jolla, CA 92093-0225, USA.}{\auorc[https://orcid.org/]{0000-0003-0112-8412}{(AG)}}

\affil[7]{Geophysics, Department of Earth and Environmental Sciences
Ludwig-Maximilians-Universität (LMU) München Theresienstraße 41 80333 Munich, Germany}{\auorc[https://orcid.org/]{0000-0003-0112-8412}{(AG)}\auorc[https://orcid.org/]{0009-0004-7825-9458}{(NS)} \auorc[https://orcid.org/]{0000-0003-4213-3322}{(ZN)}}

\affil[8]{United States Geological Survey, NASA Ames Research Park, Building 19, Mountain View, CA 94035}{\auorc[https://orcid.org/]{0000-0001-5938-6743}{(TS)}}

\affil[9]{Chair of Scientific Computing in Computer Science,
Department of Computer Science, Technical University of Munich (TUM) Boltzmannstraße 3 85748 Garching, Germany}{(SW)}

\affil[2]{Geophysical Institute, University of Alaska Fairbanks, 2156 Koyukuk Drive, Fairbanks, AK 99775 USA}{\auorc[https://orcid.org/]{0000-0003-2804-713}{(CT)}\auorc[https://orcid.org/]{0000-0001-9982-3768}{(JT)} \auorc[https://orcid.org/]{0000-0002-3901-4755}{(BC)}\auorc[https://orcid.org/]{0000-0002-2776-6431}{(AG)}}

\affil[5]{Lamont-Doherty Earth Observatory, Columbia University, Palisades, USA}
{\auorc[https://orcid.org/]{0000-0002-1286-9737}{(FW)} \auorc[https://orcid.org/]{0000-0002-1046-4525}{(KW)}  }

\affil[10]{Department of Earth and Planetary Sciences, Harvard University, 20 Oxford Street, Cambridge, MA 02138 USA} {\auorc[https://orcid.org/]{0000-0002-1877-5539}]{(CY)}}

\affil[11]{School of Geosciences, University of South Florida, 4202 E. Fowler Avenue Tampa, FL 33620, USA 813-974-2011, USA} {\auorc[https://orcid.org/]{0000-0002-8456-8001}]{(FRC)} \auorc[https://orcid.org/]{0000-0002-8541-6560}]{(JB)}}

\affil[12]{Natural Resources Canada, 601 Booth Street, Ottawa, Ontario, K1A 0E8, Canada}{\auorc[https://orcid.org/]{0000-0002-3047-9556}{(LD)}}

\affil[13]{Department of Earth and Atmospheric Sciences, Indiana University, 1001 East 10th Street, Bloomington, IN 47405-1405, USA}{\auorc[https://orcid.org/]{0000-0003-2128-9897}{(GP)}}

\corau{*Corresponding author: mdenolle@uw.edu}


\begin{abstract}
With the rise of data volume and computing power, seismological research requires more advanced skills in data processing, numerical methods, and parallel computing. We present the experience of conducting training workshops over various forms of delivery to support the adoption of large-scale High-Performance Computing and Cloud computing to advance seismological research. The seismological foci were on earthquake source parameter estimation in catalogs, forward and adjoint wavefield simulations in 2 and 3 dimensions at local, regional, and global scales, earthquake dynamics, ambient noise seismology, and machine learning. This contribution describes the series of workshops that were delivered as part of research projects, the learning outcomes of the participants, and lessons learned by the instructors. Our curriculum was grounded on open and reproducible science, large-scale scientific computing and data mining, and computing infrastructure (access and usage) for HPC and the cloud. We also describe the types of teaching materials that have proven beneficial to the instruction and the sustainability of the program. We propose guidelines to deliver future workshops on these topics.
\end{abstract}

\maketitle

\section{Introduction}
Seismological research is advancing rapidly with the rise of computational power and big data, similar to other branches of geosciences \citep{Morra-Eos2021}. Seismological research encompasses a vast range of scientific inquiries and methodological practices. Driven by often sparse but fundamental observations of earthquake phenomena at all spatial and temporal scales, seismological research has historically relied mostly on first-principle theories that supported observations. Higher education in earthquake sciences builds on this rich legacy. Most undergraduate and graduate curricula are centered around foundational textbooks, such as ``Introduction to Seismology” by \citet{shearer2019introduction}, ``Introduction to Seismology, Earthquakes, and Earth Structure" by \citet{stein2009introduction}, or advanced seismological theory, such as ``Quantitative Seismology” by \citet{aki2002quantitative}. These theoretical foundations for seismological research are typically taught in class lecture settings. 

Numerical methods and the rise of high-performance computing have fueled the development of computational seismology, notably to solve the wave equation in complex media \citep[e.g.,][]{KomaVilotte1998,Koma2002,bao1998large,olsen1996three,graves1998three} and coupled to complex source models for purposes of physics-based ground-motion simulations \citep[e.g.,][]{Graves2011} and for seismic imaging \citep[e.g.,][]{LiuGu2012,Tromp2020}. As examples, the open-source SPECFEM package (e.g., \citet{KomaTromp2002a,Koma2004}, \url{https://specfem.org/}) has supported a new era of passive-source (earthquake, ambient noise) full-waveform-inversion (FWI) \citep[e.g.,][]{Tape2009,DPeter2011,glad15,Chow2020} and the open-source SeisSol software (\url{https://seissol.org/}) enables realistic simulations of 3D earthquake rupture dynamics \citep[e.g.,][]{kaser2010seissol,pelties_threedimensional_2012,pelties_verification_2014,krenz_SC2021,gabriel2023,uphoff_seissol_2024}. 

Big data seismology is also vastly expanding, as continuous seismic data is recorded by more and more permanent stations worldwide, tens of thousands at the time of writing, and many more in temporary deployments. New methods emerged to include array processing \citep[e.g.,][]{rost2009improving}, ambient field (noise) seismology \citep[e.g.,][]{nakata2019seismic}, and machine learning \citep[e.g.,][]{kong2019machine,mousavi2022deep}. Discoveries of new tectonic and environmental phenomena invigorate the collection of large seismic data sets, leading to an exponential growth in data volumes and bringing our community to an era of petabyte-scale archives \citep{arrowsmith2022big}. Novel computing infrastructures such as cloud computing are particularly well suited for big data seismological research  \citep{maccarthy2020seismology,krauss2023seismology,Ni23}.

The broad adoption of open-source software in seismology based on Python \citep[e.g.,][]{obspy2010} or Julia \citep[e.g.,][]{jones2020seisio}, as well as version control hosted on GitHub, Bitbucket, and GitLab is transforming research practice and standards \citep{chue2022fair,barker2022introducing}. Scientific journals require publicly hosted repositories or software availability. The Jupyter project \citep{perez2007ipython,Pimentel19} encompasses a suite of interactive computing tools: JupyterLab, a modern, integrated development environment that unifies notebooks, code editors, and more; Jupyter Notebook, the classic, document-focused interface for interactive computing; and JupyterHub, a server that enables multi-user access to these environments.

Educational approaches responding to the rise of computational and big data seismology have mostly leveraged advanced theoretical seismology and well-established numerical methods at the graduate student level. \href{https://geodynamics.org/}{Computational Infrastructure for Geodynamics (CIG)} has established best practices for both software development and training workshops \citep{CIGbestpracticesSOFTWARE,CIGbestpracticesTRAINING}. ``Computational Seismology” by Heiner Igel \citep{Igel} and the associated Coursera course on \href{https://www.coursera.org/learn/computers-waves-simulations}{``Computers, Waves, Simulations: A Practical Introduction to Numerical Methods using Python"} (last accessed August 12, 2024) has effectively equipped STEM graduate students with the skills needed to solve the wave equation with a syllabus that blends numerical methods with seismological research problems. The textbook provides Jupyter Notebooks, is entirely open source in Python, and can be run for simple problems from the associated Binder hub \citep{seismolive}. Despite this, we see a growing gap between higher education curricula and research practice. Open science and novel cyberinfrastructure present opportunities to train students and researchers in current research practices. 

\begin{figure}
    \centering
    \includegraphics[width=\linewidth]{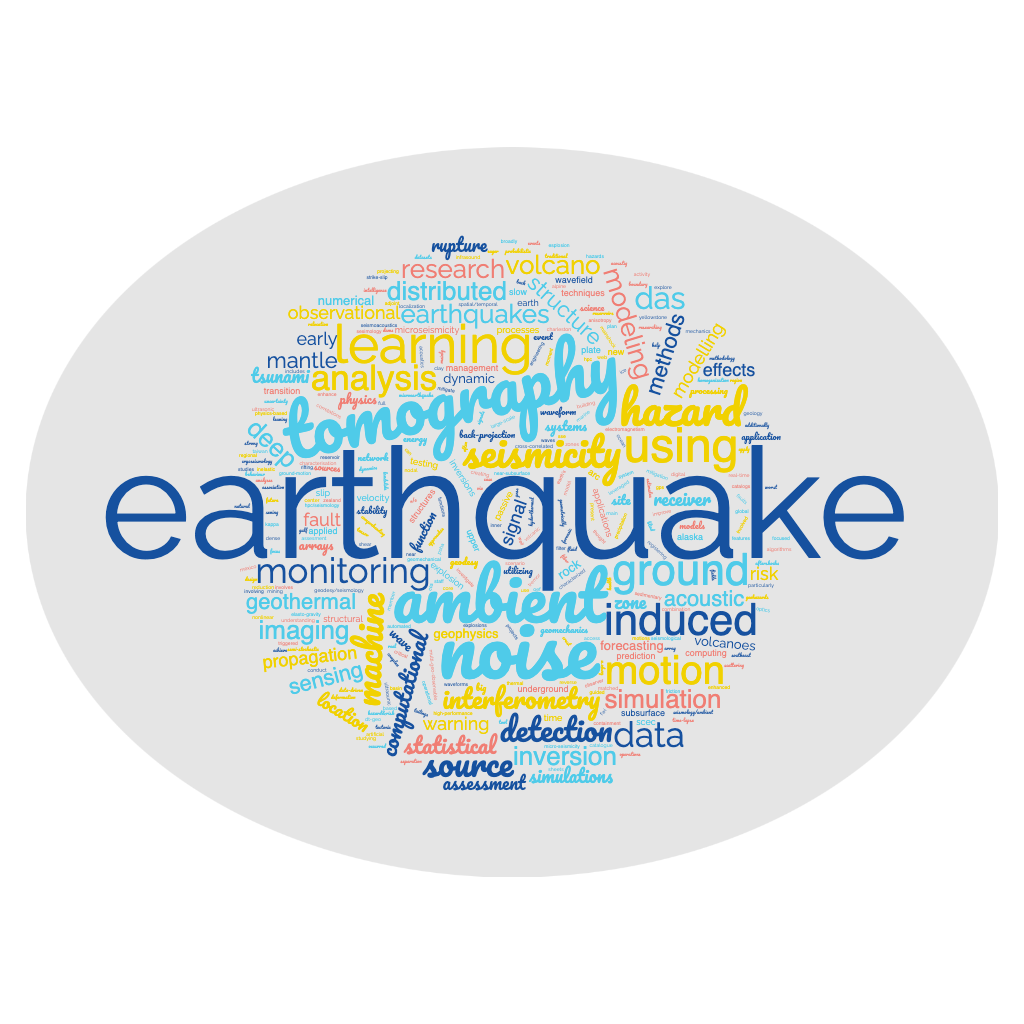}
    \caption{Word cloud illustrating the participant-reported research areas of the 2023 CyberTraining workshop.}
    \label{fig:worldcloud}
\end{figure}

The COVID-19 pandemic has transformed education: students and teachers had to transition from in-person to remote, online learning. Several efforts have contributed to improving remote access to seismology education, such as the ROSES program \citep{brudzinski2021learning}. These contributions have democratized education through pedagogical approaches analyzing small datasets, using approximate solutions, or performing modest simulations using single nodes and Jupyter Notebooks. However, a gap remains in the adoption of advanced computing platforms, such as high-performance computing (HPC) infrastructure and cloud computing. This article presents recent developments by the project \href{https://seisscoped.org}{SCOPED} (Seismological COmputational Platform to Empower Discovery  \citet{scoped2022NSFposter,Wang2023,Denolle2024}) and collaborations with other projects (e.g., \href{https://www.scec.org/}{the Statewide California Earthquake Center (SCEC)}, \href{https://www.earthscope.org/}{EarthScope}, and the European projects \href{https://www.geo-inquire.eu/}{Geo-Inquire}, \href{https://dtgeo.eu/}{DT-GEO} and \href{https://cheese2.eu/}{ChEESE-2P} \citep{Folch2023}) to help close that gap for students and researchers with multi-modal educational efforts.

The goal of the \href{https://seisscoped.org}{SCOPED} project is to develop a cyber-infrastructure that enables hybrid model--data research in seismology by utilizing cloud and HPC infrastructures, open-source software, and containerization. Software containerization is a lightweight virtualization of software and its dependencies into a portable, isolated environment, ensuring consistency across different computing environments. Research enabled by \href{https://seisscoped.org}{SCOPED} includes 1) machine-learning-enhanced earthquake source characterization and catalog building, 2) full-waveform inversion for source mechanisms, 3) full-waveform inversion for Earth imaging across scales, and 4) time-lapse imaging of the subsurface. The \href{https://seisscoped.org}{SCOPED} community expressed their research interest, which we illustrate with Fig.~\ref{fig:worldcloud}. This article discusses the workshops held as part of the \href{https://seisscoped.org}{SCOPED} project (Table~\ref{tab:workshop}), and in particular, by its use of containers. The main goal was for workshop participants to learn about research software and how to access and use high-performance computing resources, clusters from HPC centers and resources from Cloud.

\begin{table}[]
    \centering
    \begin{tabular}{|l|c|c|c|}
    \hline
        {\bf Name} & {\bf Date} & {\bf Attendance} & {\bf Range of}  \\
          & {\bf(mo/yr) }& {\bf mode}  & {\bf participant} \\
          & &  & {\bf attendance} \\
         \hline
         MTUQ & 04/2022 & Virtual & 77 \\
         SPECFEM  & 10/2022 & Virtual & 50-183 \\
          Users &  &  &  \\
         SPECFEM  & 10/2022 & Hybrid  & $\sim$ 30\\
          Developers & &   & \\
         HPS & 04/2023 & Virtual &30-80 \\
          CyberTraining &  &  & \\
         SSA & 04/2024 & In-person & 80 \\
         SCOPED & 05/2024 & Hybrid & 100 \\ 
         MsPASS & 06/2024 & Virtual & 54  \\ \hline
    \end{tabular}
    \caption{Dates and Attendance modes of the workshops. MTUQ stands for Moment Tensor estimates and Uncertainty Quantification from broadband seismic data. SPECFEM stands for SPECtral Finite Element Method. High-Performance Seismology (HPS) cybertraining. SSA stands for the Seismological Society of America. MsPASS stands for Massive Parallel Analysis System for Seismology. }
    \label{tab:workshop}
\end{table}

\section{A Broad Survey of the Seismology Community}
As part of SCOPED, we ran multiple surveys to gauge the community interest in using widely employed seismological software in advanced computing environments such as HPC and the Cloud before the workshops (see Table~\ref{tab:workshop}.) The workshops were delivered in virtual, hybrid, and in-person formats with a target audience of researchers with a skill level of advanced graduate levels in seismology. Surveys and workshops were announced in domain-specific mailing lists, such as the EarthScope Consortium and SCEC, and social media platforms (X, formerly Twitter and LinkedIn). The surveys were tailored to each workshop, and the data collected mainly focused on familiarity with the required technical skills. Overall, we received 976 responses, although some may come from the same individuals. Our workshops had a total of 574 participants, with over 130 joining in person. Some of the surveys presented here have a broad community reach, and our post-event surveys only had the workshop participants.
We found that the timing and frequency of surveys had an effect on the response rate. Post-event surveys were successful only if participants completed them during the event. Due to differences in response rates for pre- and post-event surveys, our analysis combines common questions and categories from both types. The response rate was above 96\% for requests during the workshop, whether the meeting was in person or virtual, while it was 13\% in the case of the 2024 SSA workshop.

Survey questions were designed to minimize the imposter syndrome as suggested by \cite{Huppenkothen18}. For instance, we asked participants about their familiarity with shell scripting in various forms: ``How familiar are you with computing programming from a command line (i.e., within a terminal window)?" with the response fields of ``No experience, Some Experience, Extensive Experience." We also asked about their familiarity with version control with questions such as ``All of my active research projects over the past year are on GitHub with many check-ins". Another example to assess their proficiency in Python was ``I use Python in my life" with the multiple-choice answer ``several hours a week and mostly in the classroom", ``several hours a day in my research", ``all and every day!", and ``Never-ever". We also gathered preliminary knowledge about the technical skill levels of the survey respondents. We emphasize that our surveys were a ``self-assessment," which likely provided a biased response. 

\begin{figure}
    \centering
    \includegraphics[width=0.75\linewidth]{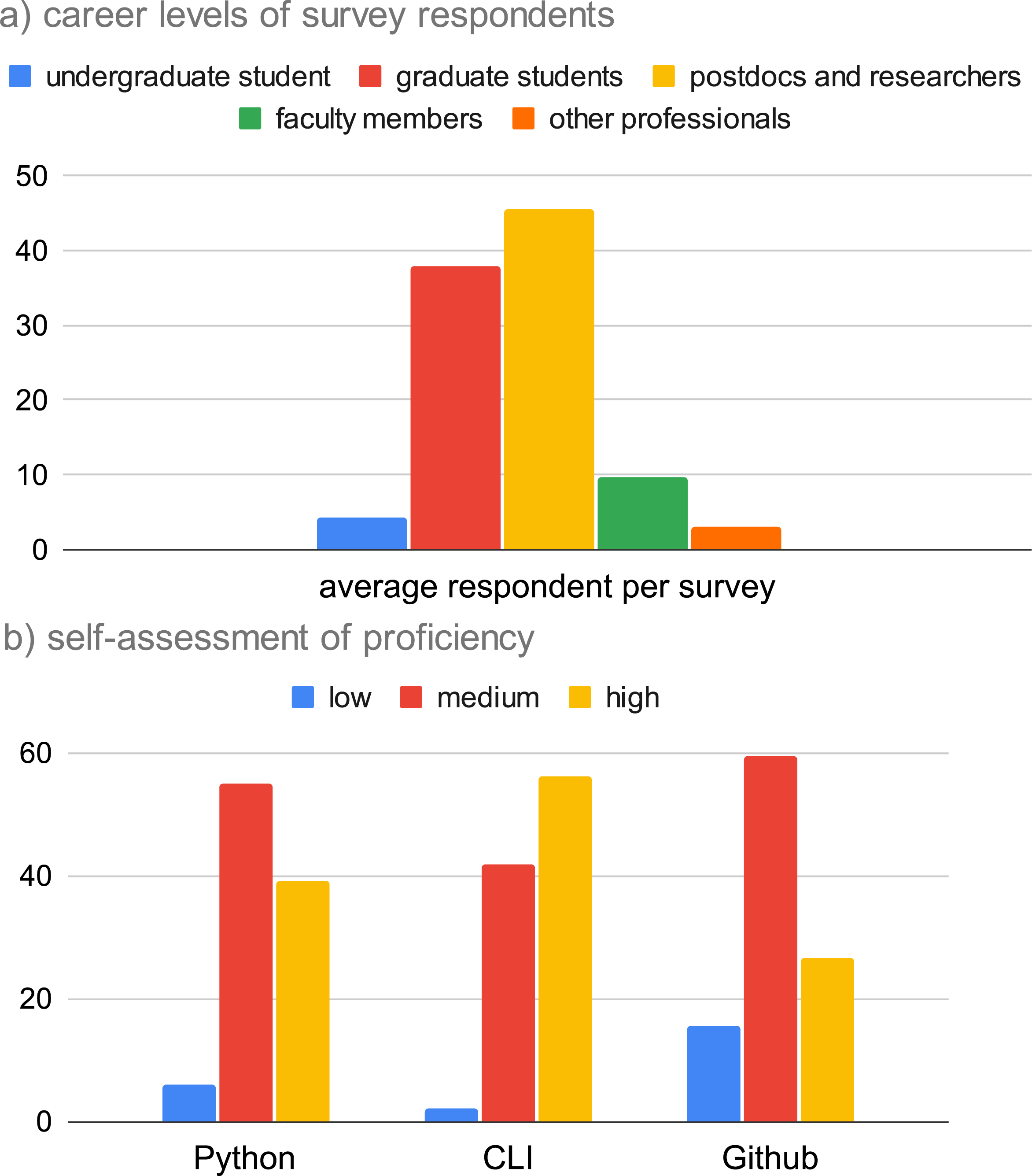}
    \caption{Proportion in percent (\%) of participants as a summary over 7 workshop surveys for a) career levels and b) self-assessed proficiency in computing tools (CLI=command line interface).}
    \label{fig:summary}
\end{figure}

Our surveys canvassed career levels of interested workshop participants, which is illustrated in Figure~\ref{fig:summary}a. The surveys included multiple choice questions with various career levels and sometimes received multiple answers. For instance, participants responded to both ``graduate student" and ``research scientist" or added an additional category of ``PhD candidate". While some surveys distinguished between ``postdoctoral researcher" and ``research scientist", we have grouped these two categories as they both represent researchers with advanced technical skills and dedicated project-based research. Out of the 976 survey responses, the demographic of the surveyed community exhibited a great majority of graduate students and researchers, with a significant participation (15\%) of faculty members. Undergraduate students have a distinctly lower participation level, likely due to our choice of communication channels and the required technical skills advertised in the announcements.

We use three skill-based proficiency levels as a relative metric for the seismological community readiness for working with our large-scale software in Figure~\ref{fig:summary}b. Overall, most participants felt comfortable with command-line interface (CLI) tools. 90\% of the participants reported having a sufficient level of familiarity with running Jupyter Notebooks and using Docker images. 40\% of the participants declared being sufficiently experts in Python for their research, though a majority declared having a medium level of familiarity. Interestingly, version control using Git ranks last in our assessment, as most participants report having a medium level of comfort, and 16\% declare having no experience with GitHub. 

\section{Novel CyberInfrastructure (CI)}
\subsection{CI on HPC}

The primary computing environment benefiting large-scale seismology today is HPC, which is enabled by clusters of thousands of tightly connected nodes managed by large computing centers, such as the resources used in our workshops, the \href{https://tacc.utexas.edu/}{Texas Advanced Computing Center (TACC)} and the \href{https://www.sdsc.edu/}{San Diego Supercomputing Center (SDSC)}. Clusters are designed for parallelized workflows. Many seismological applications, such as full-waveform inversion, seismic imaging, and earthquake cycle simulations, follow a Single-Program Multiple Data model, where the same code runs across multiple nodes, each processing a different subset of the data. These HPC workflows require optimized software with efficient parallel scaling, as well as proficiency in job scheduling, memory management, and storage architecture. HPC centers often provide training to help researchers develop and optimize these computational techniques.

Over the course of the workshops, we have trained participants in various aspects of HPC. The lectures entailed training on the fundamentals of HPC, how to write allocation proposals for HPC resources (e.g., what are the elements to show in a proposal to an NSF access proposal, or how to get AWS education or cloudbank), and the parallelization of workflows leveraging shared or distributed memory architectures. We also trained a few selected groups of participants, approximately 80 total, to access and run forward and adjoint simulations to compute 3D synthetic seismograms and data sensitivity kernels with SPECFEM3D\_GLOBE \citep{KomaTromp2002a, KomaTromp2002b} for FWI and dynamic rupture simulations with SeiSol \citep{kaser2010seissol} on the Frontera system \citep{Frontera2020} at the Texas Advanced Computing Center (TACC).

\subsection{CI on Cloud Computing}
Cloud computing is a new paradigm for computing, where users rent hardware from commercial computing centers such as \href{https://aws.amazon.com}{Amazon Web Services (AWS)}, or \href{https://azure.microsoft.com}{Microsoft Azure}, which provide on-demand and a-la-carte hardware choices. Computing is done on ``virtual machines" (VM), an abstraction of hardware that contains up to a few hundred CPU cores, up to a few GPUs, and a tunable amount of memory. Maximum-size instances can have up to about 200 cores, 10 GPUs, and 1TB of memory and are designed mostly for big-data processing, for example, when training complex machine-learning models. VMs have a pre-loaded operating system on which users install dependencies from scratch, from Docker images, or from previously saved virtual images.

Cloud computing is still in its infancy in seismology, and user access remains a challenge \citep{krauss2023seismology}. 
We trained participants in cloud computing concepts, such as its design to interact with storage and perform large-scale deployments. We presented diverse strategies for using cloud resources to workshop participants. They accessed Google Colab (\url{https://colab.research.google.com/}) provided by Google Cloud Platform, which is a pre-configured Python-based Jupyter Notebook, and learned how to customize them by manually installing additional dependencies. Accessibility is a major benefit of the Colab approach, as VM specifications can easily be modified on the Google Colab web interface. The free version of Google Colab is limited in size (e.g., a few CPUs, 12 GB of RAM, and 50 GBs of storage).

\begin{figure}
    \centering
    \includegraphics[width=\linewidth]{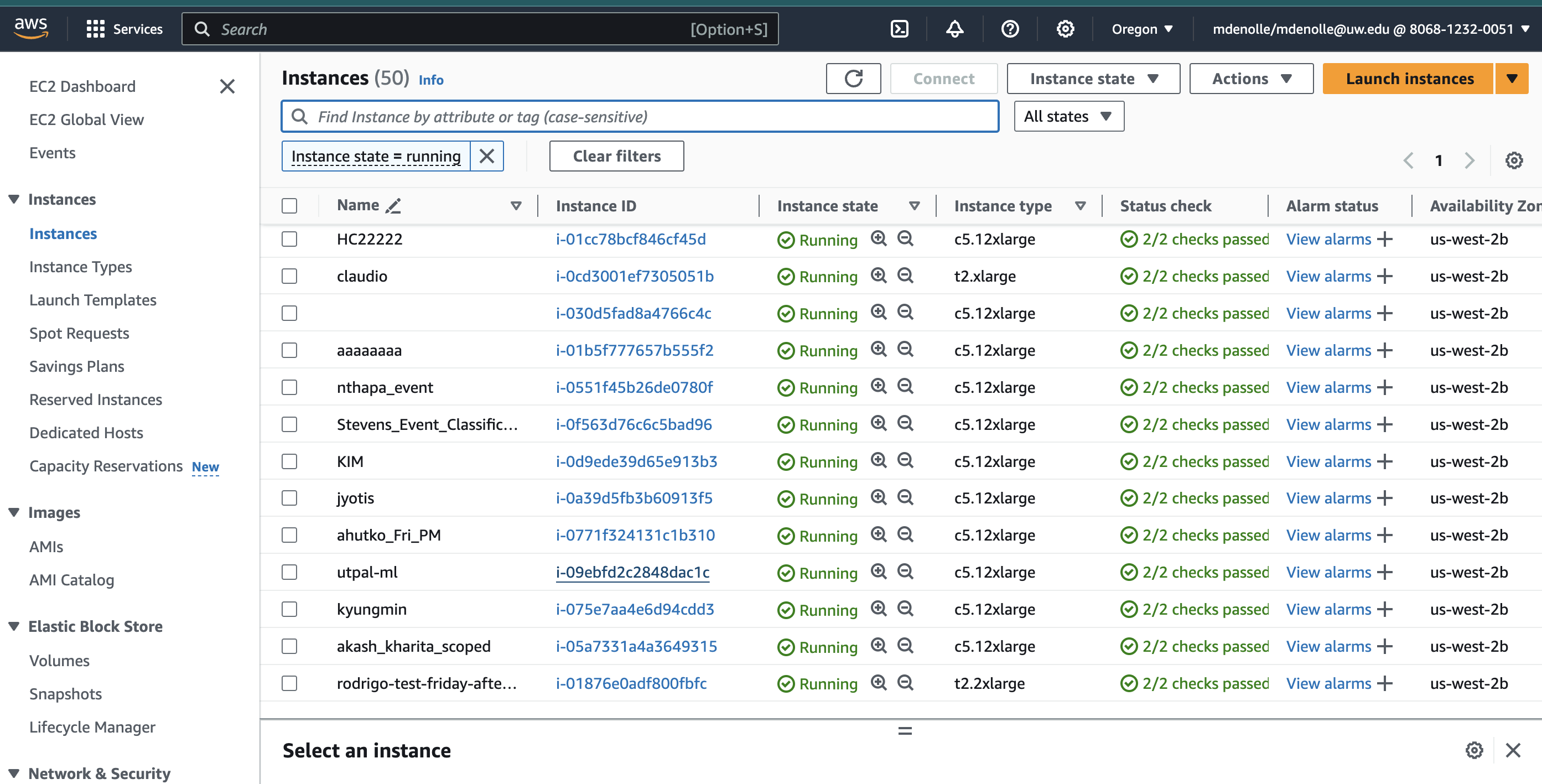}
    \caption{Web browser screenshot showing AWS instances created by workshop participants during the 2024 SCOPED workshop, illustrating multiple instances running simultaneously on the same allocation. Instance participants chose names.}
    \label{fig:instances}
\end{figure}

The SCOPED project chose AWS as the cloud provider due to the availability of large seismic datasets already hosted on AWS Simple Storage Service (S3) (Northern and Southern California data centers as NCEDC and SCEDC \citet{NCEDC, yu21scedc}). The workshop covered 1) various ways to access AWS cloud resources, 2) how to launch an AWS computing resource on the Elastic-Computing (EC2) referred to as an {\it instance} from scratch via the web console, 3) how to install basic research software into their instances, and 4) how to run research-grade problems on the Cloud. We used the typical AWS web console to deploy compute resources during one of our workshops and illustrated it in Figure~\ref{fig:instances}. We taught basic concepts and practiced popular tools for software environments and versioning, such as {\tt git}, {\tt Docker}, and {\tt conda}. We note that significant effort was required to simplify and prepare instructions for streamlined access and use of AWS instances. In particular, it is not trivial to open and access a Jupyter Lab, and we curated the training materials to achieve this in our Jupyter Book (\href{https://seisscoped.org/HPS-book/intro.html}{HPS; High Performance Seismology \citep{HPS2024}}). 

Additionally, we taught various ways to conduct research workflows on the cloud: cloud-native workflows that incorporate cloud services as part of the design (e.g., NoisePy, \cite{jiang2020noisepy}, \href{https://seisscoped.org/HPS-book/intro.html}{HPS}), and, alternatively, workflows that are lifted-and-shifted migrated to the cloud, e.g the Lamont-Doherty Earth Observatory earthquake catalog production workflow \citep{Wang24} that includes algorithms for event detection and phase arrival time measurements (QuakeFlow, \cite{zhu2023quakeflow}), discrimination (SpecUFEx, \cite{holtzman2018,sawi2022} ), and relocation (HypoDD, \cite{WaldhauserEllsworth2000}).

\subsection{Open-Source and Containerized Software}

The SCOPED platform gathers open-source software that tackles big data and large-scale software research. Currently, SCOPED includes full waveform modeling and inversion, machine-learning-aided earthquake catalog building and source characterization, ambient field seismology, and earthquake dynamic rupture simulations. 

The underpinning strategy for deploying our software is containerization, which enhances portability and exploits negligible computing overhead \citep{Wang_container} once successfully containerized. Containers are isolated images of software and its dependencies that can be deployed on various operating systems and hardware (\href{https://www.docker.com}{Docker}, {Singularity}). Containers promote long-term sustainability and reproducibility of the computing analysis. To grow our user and developer community, SCOPED flagship software is containerized with tutorials provided in the form of Jupyter Notebooks. We developed a \href{https://github.com/orgs/SeisSCOPED/packages}{SeisSCOPED container registry} in which the container base holds minimum dependencies. Additional dependencies can be added to the container base: for instance, an HPC-specific container loads modules for Message-Processing Interface - MPI, and a cloud-specific container has cloud-provider Command-Line-Interface CLI-specific packages. One significant advantage of the containerized software approach is long-term stability; workshop users and future students alike can leverage the same workshop container and its pinned software dependencies, training materials, and test data. Another powerful use of notebooks is integrating shell scripting within notebook cells using system commands. As an example, one can deploy parallelized Python scripts on Azure Pool \citep{krauss2023seismology}, AWS Batch resources in a single notebook, or run parallelized SPECFEM simulations through Jupyter Lab  (\href{https://seisscoped.org/HPS-book/intro.html}{HPS}).

While high-performance computing favors the use of scripting, compiled executables, and minimal container sizes, training materials benefit from attaching small test data and notebooks for documentation and visualization of results. Containers may add {\tt IPython} and {\tt ipykernel} dependencies to support Jupyter Notebooks and small test data to a given container. Especially for cloud computing, opening Jupyter Notebooks from remote servers can pose a challenge in group settings, especially on cloud instances. Throughout workshops, our team came up with the following command line to easily allow access to Jupyter Notebooks from a container by fixing the token and IP address:
\begin{lstlisting}[style=pythonStyle]
sudo docker run -p 80:8888 --rm -it ghcr.io/seisscoped/noisepy:centos7_jupyterlab\
    nohup jupyter lab --no-browser --ip=0.0.0.0 --allow-root  --IdentityProvider.token=scoped &
\end{lstlisting}
where {\tt IdentityProvider.token=scoped} gives a specific token (this avoids users tracking it in the long logs printed on the terminal), {\tt allow-root} grants root access for users inside the container volume, {\tt ip=0.0.0.0} tells the server to listen on all available network interfaces, and {\tt nohup jupyter lab --no-browser \&
} starts Jupyter Lab in the background without attempting to open a browser because launching a browser is not possible on a remote virtual machine that lacks a graphical user interface and is protected by a private IP address. This small code snippet was designed to accelerate research rather than being hung up on infrastructure.

\subsection{Open Education}

\begin{figure*}
    \centering
    \includegraphics[width=\linewidth]{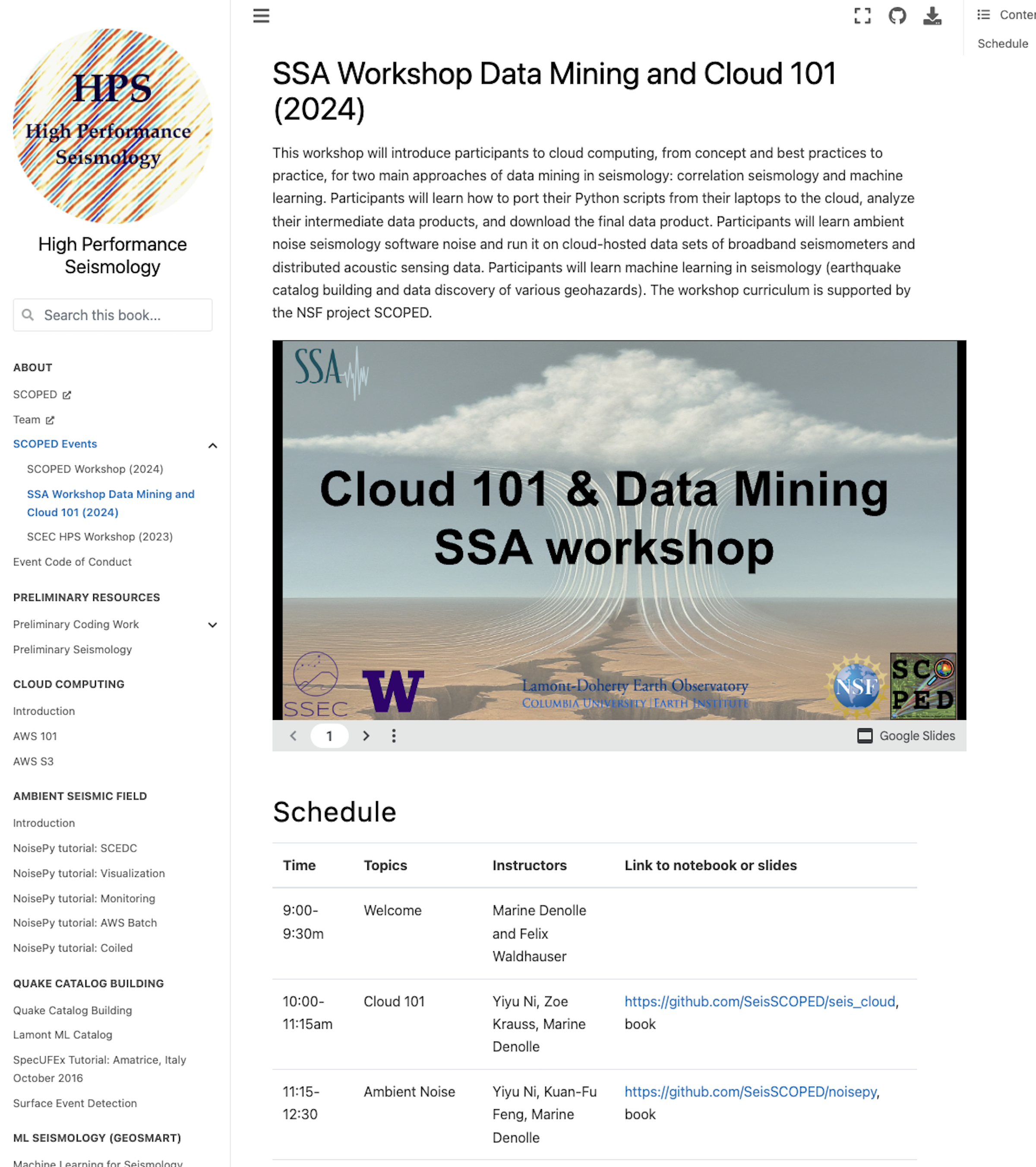}
    \caption{A page of the HPS Jupyter Book for the SSA 2024 workshop, which embeds a Google slide presentation for the introduction presentation, the schedule of the specific workshop, and links to relevant book pages}
    \label{fig:hps-ssa}
\end{figure*}

Open Education, a set of practices and principles aimed at making learning opportunities more accessible and equitable for everyone, is a promising future direction for higher education as research becomes increasingly specialized and training materials require extensive, globally distributed expertise. Jupyter Book is an appropriate platform for collaborative research education, as many instructors can contribute, and students receive up-to-date materials. Such an example is shown in Figure~\ref{fig:hps-ssa}. The challenge remains in curating training materials, as many come from complex research literature and free, non-peer-reviewed online materials. 

We are compiling a dynamic textbook titled ``High-Performance Seismology" (HPS) that the workshop instructors have contributed to \href{https://seisscoped.org/HPS-book/intro.html}{HPS} \citep{HPS2024}.

\section{SCOPED-related events}
\subsection{Virtual Events}

We have conducted several virtual events, which offer great potential for democratizing access to advanced computing globally. To maximize participation, we structured events into short sessions ($\sim$ 45 min) with adequate breaks and scheduled them at times that accommodate participants across various time zones. Pre-event surveys of user locations helped select optimal event times, ensuring broad participation. Additionally, we recorded the training events and made them asynchronously available on our \href{https://www.youtube.com/@scoped6259}{SCOPED Youtube channel} to address time zone conflicts. Over the channel's lifetime and until February 16, 2025, it has seen 2500 views with 250 hours of course content.

\begin{figure*}
    \centering
    \includegraphics[width=\linewidth]{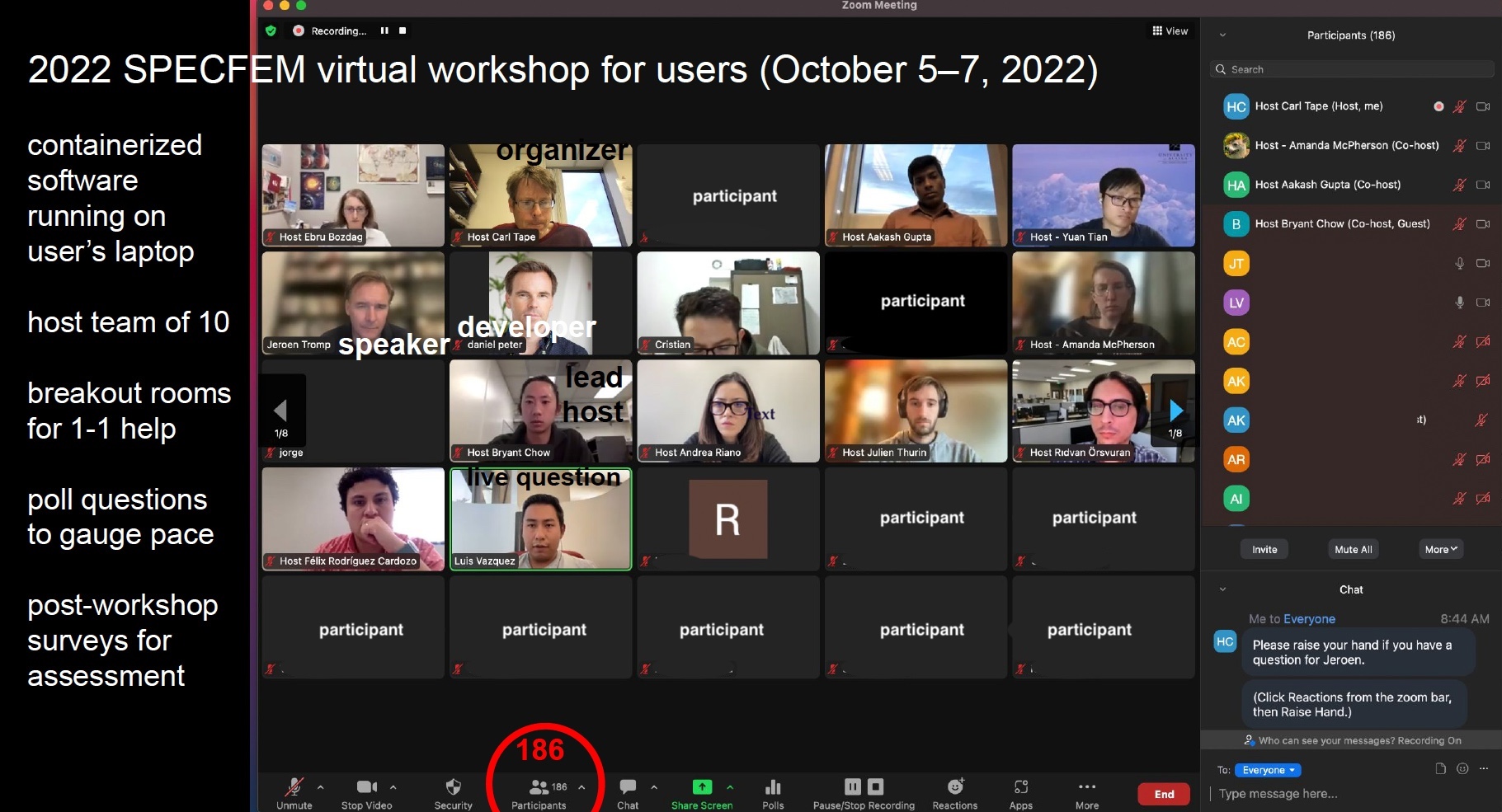}
    \caption{Annotated zoom screenshot from the SPECFEM virtual workshop for users (October 5, 2022). At this stage in the workshop, there were 186 participants (red circle). The annotations and windows show the hosts, the speaker, the lead software developer, the lead host/instructor, the organizer, and a participant asking a question.}
    \label{fig:specfem-zoom}
\end{figure*}

In April 2022, we organized a two-day workshop on moment tensor estimation using the open-source MTUQ software \citep{Thurin2023SSAabstract}. The first day featured a 2-hour session introducing key concepts and tutorials. The second day consisted of a 4-hour session that demonstrated how to calculate a library of Green's functions for a specified 1D layered model using a frequency-wavenumber code \citep{ZhuRivera2002} and obtain a seismic moment tensor solution. Attendance was strong, with 78 on day 1 and 68 on day 2, indicating sustained interest in the more detailed content. In preparation, software containers for four systems (Windows/PC, Linux, Mac OS Intel, and Mac OS Apple Silicon processors) were developed and tested, resulting in high success rates for participants running the examples.

Building on the success of the previous workshop, we held a three-day SPECFEM users' workshop in October 2022. Each of the three daily, 4-hour sessions had a specific focus: the forward wavefield (day 1), sensitivity kernels (day 2), and seismic imaging (day 3). Each session included short (20-minute) science lectures, 45-minute tutorials that participants could run locally using pre-downloaded software containers, and wrap-up discussion sessions. Participation ranged from  187 attendees in the day~1 opening seminar to 63  in the day~3 discussion (Figure~\ref{fig:specfem-zoom}). 

This was the first SPECFEM workshop featuring seismic imaging, providing a natural progression from synthetic seismograms (day~1) to sensitivity kernels (day~2) to iterative tomographic inversion using SeisFlows and Pyatoa (day~3) \citep{seisflows,Chow2020}. Crafting a pedagogical but research-grade notebook took a dedicated effort, given the differences in objectives (performance and robustness vs clarity and interactivity). We took some steps to exemplify but downsize large-scale processes in order to retain the same outcome. On the big-data analysis, this meant choosing the duration of the experiment (e.g., 1 day of data) or the spatial extent (e.g., number of stations) to be feasible on 2-4GB worth of RAM, but using the same software tools for the 100 times bigger scale.

In 2023, we held a four-day virtual training workshop in collaboration with the SCEC and several European and NSF-funded projects. Each day focused on a specific theme, starting with an opening day of lectures on open science, reproducibility, software best practices, and an introduction to HPC and Cloud Computing. Subsequent days were divided into subdisciplines and platforms. The workshop attracted over 200 interested participants, with 80 joining on Zoom at the workshop's start, though attendance varied due to time zone challenges. As the workshop was designed for tool adoptions and relatively fast-paced, participants were exposed to diverse topics in seismology, including earthquake simulations focusing on dynamic rupture (with SeisSol, \cite{kaser2010seissol, uphoff_seissol_2024}) and wave propagation (SPECFEM, \cite{Koma2002}), machine learning phase picking (ELEP, \cite{yuan2023better}), earthquake probabilistic forecasting (pyCSEP, \cite{savran2022pycsep}), and user access to Community Earth Models maintained by SCEC, \citep[e.g.,][]{Plesch2007,Small2017}. 

The 2024 MsPASS training short course was hosted in collaboration with EarthScope during the week of July 8, 2024, as part of their 2024 Technical Short Course series. The event featured two hours of lectures and hands-on sessions over three days. Participants could attend the course in real-time or access recordings on YouTube afterward. Daily homework assignments were given, and an optional final project. The application-based enrollment process received 99 valid applications, from which 53 participants were accepted to attend. The cohort was notably diverse, with 38\% self-identifying as underrepresented in the geoscience community and 26\% identifying as female. This short course was the first event on EarthScope's \href{https://www.earthscope.org/data/geolab/}{GeoLab} platform, a new experimental cloud-based Jupyter Lab platform hosted by EarthScope. MsPASS \citep{wang_mspass_2022} was the first application to run parallel processing workflows on the GeoLab platform. Participants were exposed to topics such as using MsPASS to process waveform data in the cloud, managing datasets with a document database, and executing data processing workflows in parallel. 6 participants completed all of the homework assignments, 4 had partial submissions, and 1 completed the optional project.

\subsection{In-person Events}
We held a one-day workshop at the SSA meeting in Anchorage in 2024 that was in-person only, a fast-paced event with an introduction to cloud computing and research workflow. We successfully had 80 participants launch their own cloud instances on AWS, where they detected earthquakes in cloud-hosted SCEDC data and output data products to a shared MongoDB database. Participants also ran machine learning workflows for earthquake catalog building, including supervised and unsupervised learning approaches for event-type classification. The SSA participants enrolled as a first-come, first-serve approach, and communication with the participants was not as well-established as in the other SCOPED events. This slowed down the initial steps of setups, and advanced participants had to follow the pace of beginners. 

\begin{figure*}
    \centering
    \includegraphics[width=\linewidth]{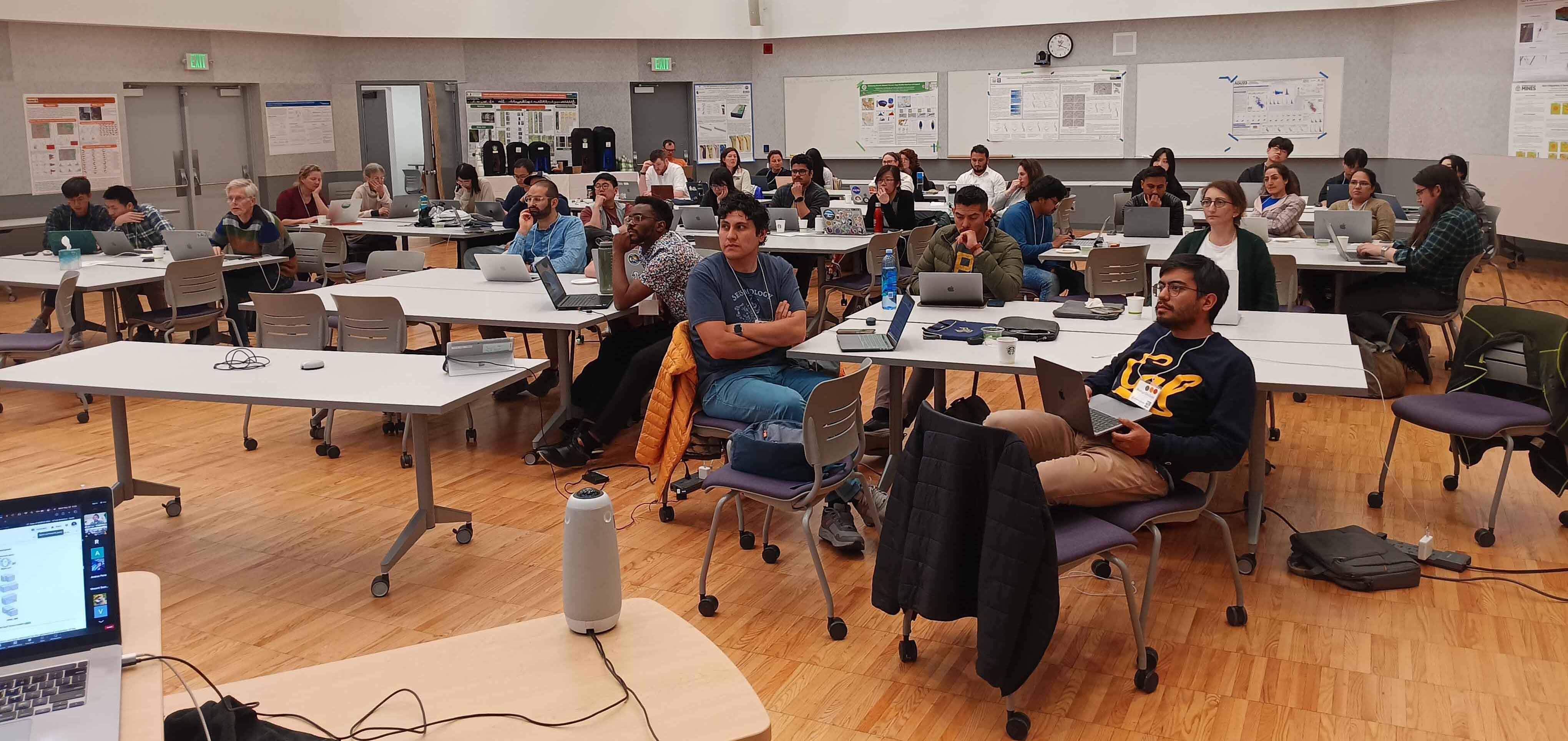}
    \caption{In-person component of the 2024 SCOPED workshop. Participants engaged in live exercises. The OWL camera and directional microphone (bottom center), together with Zoom (speaker's laptop at lower left), enabled hybrid participation. Participants' posters can be seen on the walls.}
    \label{fig:scoped2024}
\end{figure*}

Our most recent SCOPED workshop was a five-day hybrid meeting at the University of Washington in Seattle in May 2024 (\href{https://seisscoped.org/workshop-2024/}{https://seisscoped.org/workshop-2024/}, last accessed 09/10/2024). About 50 participants, including instructors, attended the workshop in person (example of room layout in Fig.~\ref{fig:scoped2024}), along with a varying number of online participants (on average about 50 per day). The training program was led by the research groups of the SCOPED PIs and Dr. Alice-Agnes Gabriel from the University of California San Diego, supported by several NSF- and European-funded projects. Day 1 covered subjects and practicals with an introduction to HPC and Cloud computing and best practices for developing and maintaining open-source software. Day 2 was dedicated to 2D and 3D wave simulations with SPECFEM packages and a tutorial on introduction to full-waveform inversion by SeisFlows and moment tensor inversions with the MTUQ software. Day 3 focused on 3D dynamic rupture and finite source earthquake simulations with SeisSol and 3D wave simulations and computation of 3D adjoint data sensitivity kernels \citep{Tromp2005, bozdaug2011misfit} for full-waveform inversion on a one-chunk mesh with SPECFEM3D\_GLOBE \citep{KomaTromp2002a, KomaTromp2002b}. Day 4 focused on high-precision earthquake catalogs \citep{Wang2024}, where they combined machine learning algorithms (QuakeFlow; \cite{zhu2023quakeflow}) with large scale cross-correlation and double-difference methods (HypoDD; \cite{WaldhauserEllsworth2000,waldhauser2008large} and demonstrated the use of unsupervised machine learning (SpecUFEx;  \cite{holtzman2018, sawi2022}). The participants also had a session on MsPASS \citep{wang_mspass_2022} to learn how to manage big data on HPC and the Cloud. Day 5 addressed ambient noise seismology on the cloud (NoisePy; \cite{jiang2020noisepy}), and the trainees were given tutorials on machine learning workflows for seismology on the cloud. All the workshop tutorials were prepared in Jupyter Notebooks hosted on GitHub, containerized versions of the open-source SCOPED software were used, and lectures were recorded and uploaded to the \href{www.youtube.com/@scoped6259}{SCOPED YouTube channel}. All 3D simulations on Day 3 were performed by the trainees on the Frontera system, and the observational seismology tutorials were on AWS.

\subsection{Learning Objectives and Outcomes}
Our main learning objectives were for participants to 1) be able to explain the fundamental principles of high-performance and cloud computing in seismology, 2) apply appropriate computing resources (e.g., HPC clusters, AWS instances) to execute research workflows, 3) compare different computational strategies for seismological research (e.g., traditional local computing vs. cloud/HPC-based approaches), and 4) evaluate their efficiency in handling large-scale seismological data.
Surveys following the SSA and HPS CyberTraining workshops enabled us to evaluate some of these learning outcomes. In the SSA survey, we evaluated the learning outcomes of each module, which were about cloud computing and research-grade applications in ambient noise seismology and machine learning in earthquake catalog building. Eleven survey respondents out of eighty participants noted improved cloud computing skills and overall self-reported positive learning outcomes with the workshop. Future improvements in workshop materials and delivery mechanisms will enhance the impacts of the training. 

In the HPS CyberTraining survey, 23 participants responded and expressed positive learning outcomes, with 70\% ranking their satisfaction 5/5 and 62.5\% indicating that the workshop was a valuable use of their time (rank 5/5). Positive learning outcomes were on Docker and reproducible \& open science, frontier seismological topics, and HPC and cloud computing. Several participants expressed verbally or via the survey that the pace was fast and that instructors should slow down when going through code blocks in notebooks, along with improving participant-led exercises in the notebooks with empty cells.

\section{A Guide to Advanced Computing Workshops}
The development of teaching materials requires dedicated effort from both faculty and participants. In-person workshops need a participant-to-assistant ratio of about 15 to 1 for effective debugging. Recruiting participants with similar technical skill levels ensure consistent progress or additional instruction time can be provided for beginners.

\subsection{Surveys}
Evaluation surveys can be helpful in quantifying learning outcomes, and crafting them with consideration can benefit professional educators and evaluators. Employing a more standardized approach to the surveys may improve their usefulness. For example, metrics such as ``None, Little, Moderate, Quite a bit, Complete'' for levels 1 through 5 are similar to the Likert 6-point, ``strongly disagree, disagree, somewhat disagree, somewhat agree, agree, strongly agree.'' More standard metrics, such as the Likert 6-point, will be incorporated in future surveys to provide a more nuanced measure of fields \citep{Huppenkothen18}. Allowing only one response per question is essential, as multiple answers can hinder post-event quantitative analysis.

\subsection{Content}
Each of the SCOPED workshops had various designs and, overall, was packed with tutorials. We developed or used several forms of pedagogy for training workshops, which individual workshops may have combined:
\begin{itemize}
    \item scientific lectures, especially those that motivate the use of advanced computing resources.
    \item lectures on cyberinfrastructure, research ethics, and software best practices.
    \item core package tutorials to train participants in using a specific software in its generic form.
    \item research-grade workflow tutorials with assisted walk-throughs.
    \item group or participant-led activity (hackathon-style).
\end{itemize}

Taken together, these pedagogies may form a module, a self-contained unit that includes 1) a brief lecture (20-30 minutes) introducing key concepts and their relevance to scientific applications, 2) a hands-on tutorial (60-90 minutes) where participants apply the concepts through structured exercises, such as running computations on cloud platforms, setting up HPC environments, or analyzing seismic datasets, and 3) a guided exploration and Q\&A (30-60 minutes) to allow participants to troubleshoot their workflows and gain deeper insights. Modules form half-day activities. In-person workshops may combine two modules per day, with 3 hour sessions striking a good balance between depth and independent exploration without too much cognitive fatigue; virtual workshops may spread these modules over multiple days and time zones to help attract foreign participants. This strategy worked particularly well for the SPECFEM and MTUQ workshops. 



\subsection{Before the event}

\subsubsection{Instructor Coordination}
Instructor coordination is critical for workshop effectiveness, which was not consistently undertaken in our workshops. We recommend {\it pre-workshop planning meetings} to define learning objectives, align instructional materials, and anticipate challenges, {\it mock run-throughs} with non-participating students or colleagues to identify unclear instructions and potential bottlenecks in execution, {\it role assignment among instructors}, enduring that each focuses on specific tasks such as concept explanation, hands-on support, or software troubleshooting, and {\it diverse instructor background} to represent different expertise area (e.g., computational scientists in HPC/Cloud, seismologists form scientific applications, software developers) to allow for a more comprehensive learning experience. 




\subsubsection{Materials \& Platforms}
We used Jupyter Books or Google Docs as shareable, open platforms to organize the workshop schedule and share training materials, with a clear first page with the schedule that links to the sources (e.g., YouTube recording, Zoom links, GitHub repositories, etc.). To-date, our teaching materials are still available and youtube videos still watched. Communication platforms like Slack or Teams, or other forms of group communication and direct messaging, allow for rapid, practical communication among instructors and between participants and organizers. Pre-workshop materials (e.g., pre-requisite tutorials, recorded lectures and videos from previous workshops, and software installation guides) help participants familiarize themselves with foundational concepts beforehand. Instructors may pre-download data, e.g., pre-processed data and static visualization, to ensure the workshop runs smoothly even with unforeseen technical issues such as loss of network connectivity.
To accommodate different learning styles, workshops may provide slides and annotated code for visual learners, interactive coding exercises for experiential learners, and may incorporate discussion-based problem-solving for verbal learners.

\subsubsection{Accounts}
We found that the workshop ran more smoothly when participants' accounts on the computing resources was set up days in advance. We provide guidance to automatically create user accounts for AWS on the  \href{https://seisscoped.org/HPS-book/intro.html}{HPS book}. It is important to remind participants that workshop computing resources are {\it temporary}. Educational allocations at HPC centers are typically provided when supercomputer center research scientists are involved in the workshop. For our workshop, users chose a simple username, for example, the participants' email address or its prefix  (e.g., <yourID>\@email.edu), as well as a single generic password to avoid manual and complicated intervention. Cloud accounts can be created at any time and managed during the workshop. For instance, a cloud manager can re-assign policies, roles, and temporary passwords during the event if needed. Through surveys and emails and possibly ``Day 0'' virtual help sessions, instructors may find it useful to ensure that computing setups (accounts, software containers) are working in advance for all virtual participants. 

Participants were made aware that these accounts were temporary and provided with guidelines on how to access these platforms in the future.

\subsection{First Day}
This is the day to onboard participants, ensure that the accounts to HPC and the Cloud are set up and accessible, install ancillary software, and download and test workshop containers to ensure they perform as expected on the participants' platforms. These tasks can also be done prior to the workshop to free up actual workshop time. Hybrid workshops can be challenging to deliver. They require multiple cameras and microphones for large rooms, attention to remote attendance, and interaction with remote participants.  For hybrid events, organizers may find it beneficial ensure that there are sufficient staff/instructors online who can help manage remote participants. Engagement can be improved with frequent polling. 

\subsection{During the Event}
The feedback on workshops has been positive, especially for focused, single-tool, and single-platform sessions. 
For virtual workshops, a helpful strategy is to ask, "Are you ready to move on?" with the options "Yes, Almost, and No." This helps pace the session and provides instant feedback on participant experience, showing engagement levels and areas needing assistance.

Some of the tutorials, especially those for the core software, included additional cells in the Jupyter Notebooks so that participants could test various parameters independently, which was implemented in SPECFEM, MTUQ, and SeisSol. Other full-stack, research-grade tutorials (e.g., ML-aided earthquake catalog building or ambient noise seismology) included advanced workflows tailored to specific use cases, making it challenging to strike a balance between teaching fundamental concepts and realistic scenarios.

Teaching cloud and HPC computing strategies for research-grade analysis can be challenging for participants if the content is not relevant to their work. In several tutorials, we chose to demonstrate how to adapt a homegrown software stack based on a specific platform to provision a cloud instance. Such an approach allows researchers to upload and deploy their own software stacks on the cloud, ensuring flexibility and independence without imposing a specific platform or software style. While this requires coordination among instructors, it enables each researcher to bring up their preferred tools, reflecting the natural workflow of scientific research.

\subsection{Post Event}
Surveying the participants is a good way to measure learning outcomes. We found that most participants will not fill out post-event surveys unless asked {\bf at the time of the events}, both for virtual and in-person meetings. To improve on the evaluation, the exit survey may benefit from having similar questions to the incoming survey. Leaving an empty box at the end permits participants to speak freely of things that worked and things to improve.

Assessments need to be more quantitative, with more structured responses than were provided in many of the surveys we ran. Some respondents provided several answers to the same question, posing additional problems in the analysis of the survey in post-processing. Further automation of the survey, such as more rigorous Python-based post-processing, will improve the reproducibility of the survey analysis. 

\section{Conclusions}

The diversity of workshops is essential to reach multiple pedagogical goals. Large attendance in virtual meetings allows for a global reach and democratization of training and access to computing resources. The size of these virtual meetings was not optimal for spontaneous communication and career network --- although future workshops could take this into account. We found that at that scale (200+ participants), it was easier to have participants run containers and software locally, whereas, for smaller, virtual meetings, it is possible to provision remote participants with temporary cloud accounts.

In-person meetings are well suited for career development, building collaborations, and provisioning participants with more advanced computing resources, which may be limited to certain countries. These in-person meetings can run longer than virtual events, with the caveat that organizers may consider pacing the delivery of the materials more slowly than they anticipate and even include participant-led hackathons for better learning outcomes and stronger cohort building.

Advanced computing with projects such as services for Jupyter Hubs (e.g., Infrastructure-as-a-Service Iaas such as  \href{https://docs.2i2c.org/}{2i2c} that support centralized servers running Jupyter Lab or Notebooks with multiple-user access), or Python projects that manage distributed cloud resources such as \href{https://www.coiled.io/}{Coiled} , and the up-and-coming science gateways \citep[e.g.,][]{airavata, HUBzero, Tapis} promote ease of access to resources, potentially benefiting the user community. Nevertheless, training the community in the concepts of cloud computing and HPC for new {\it developers} remains important so that they can continue innovating solutions for large-scale computing for seismological research and that their expertise lasts beyond the lifetime of specific Iaas. 

Our efforts in conducting these workshops reflect a positive outlook for seismologic research in the 21st century. As big seismic data become more widely accessible, seismologists at all career levels desire to pursue training in HPC and Cloud computing. We highlight the benefits of our workshop model by uniting cyberinfrastructure and research professionals skilled in HPC and Cloud computing. They leverage large-scale computing to solve seismologic problems. Through these workshops and their associated teaching materials, we are able to disseminate that collective knowledge in an open, sustainable, and reproducible manner, all to accelerate the pace of seismologic discovery.


\begin{datres}
The survey data came from Google Forms responses. Because of the lack of anonymity in the responses, the authors decided not to share the original data. All SCOPED educational materials are open-source (e.g., \url{https://seisscoped.org/HPS-book/intro.html}{HPS}). Video recordings of our workshops are available on \href{https://www.youtube.com/@scoped6259}{SCOPED YouTube channel}.
\end{datres}

\section{Declaration of Competing Interests}
The authors acknowledge that no conflicts of interest have been recorded.

\begin{ack}
Any use of trade, firm, or product names is for descriptive purposes only and does not imply endorsement by the U.S. Government. We are grateful to associate editor Alan Kafka, reviewers Brad Aagaard and Clara Yoon for their constructive feedback.
This work is supported by the Seismic Computational Platform for Empowering Discovery (SCOPED) project under the National Science Foundation (award numbers OAC-2103701 (UW), OAC-2104052(UAF), OAC-2103621 (CSM), OAC-2103741 (CU), OAC-2103494 (UT)). The events were also supported by the eScience Institute, a SCEC grant 22162. The MTUQ workshop was partly sponsored by the Geophysical Detection for Nuclear Proliferation (GDNP) University Affiliated Research Center (UARC) Task Order 7, funded by the Air Force Research Laboratory under contract HQ0034-20-F-0284. 
The HPS CyberTraining and the SCOPED training were additionally supported by the Southern California Earthquake Center and USGS (SCEC project 22162), by the National Science Foundation (MTMOD, grant no. EAR-2121568, CSA-LCCF, grant no. OAC-2139536, QUAKEWORX, grant no. OAC-2311208), the European Union's Horizon 2020 research and innovation programme (TEAR ERC Starting; grant no. 852992) and Horizon Europe (ChEESE-2P, grant no. 101093038; DT-GEO, grant no. 101058129; and Geo-INQUIRE, grant no. 101058518). Waveform data, metadata, or data products for this study were accessed through the Northern California Earthquake Data Center (NCEDC), doi:10.7932/NCEDC.

We thank the Texas Advanced Computing Center (TACC) at the University of Texas at Austin for providing computational resources on `Frontera' system \citep{Frontera2020}.

We thank Zhengtang Yang, Weiming Yang, and Jinxin Ma, who contributed to the development of \href{https://mspass.org}{MsPASS} and maintained the SCOPED containers.
We thank Dave A. May for a guest lecture on ``Open and Reproducible Science" during the HPS CyberTraining.
We thank all \href{www.seissol.org}{SeisSol} developers, specifically Thomas Ulrich, Iris Christadler, Mathilde Marchandon, Jeena Yun, Nicolas Hayek, and Jonatan Glehman, for their support in preparing and during the HPS CyberTraining and the SCOPED training. The group also thanks the SCEC and specifically Pablo Iturrieta, Jose Bayona, Phil Maechling, Fabio Silva, Mei-Hui Su, and Scott Callaghan for giving a full day of tutorials in the 2023 HPS CyberTraining workshop, tutorials that are also integrated into the HPS book. The 2024 MsPASS training short course was also sponsored by the EarthScope Consortium. We thank Sarah Wilson, Robert Weekly, Chad Trabant, Melissa Weber, Gillian Haberli, and Tammy Bravo, as well as other staff members of the EarthScope Consortium who provided technical and pedagogical support throughout the event. 

This draft manuscript is distributed solely for purposes of scientific peer review.  Its content is deliberative and predecisional, so it must not be disclosed or released by reviewers.  Because the manuscript has not yet been approved for publication by the U.S. Geological Survey (USGS), it does not represent any official USGS finding or policy
\end{ack}


\def \agu{Am.~Geophys.~Un.} \def \usgs{U.S.~Geol.~Survey} \def \dggs{Alaska Div. Geol. Geophys. Surv.} \def \antsci{Antarctic~Science} \def \aapg{Am.~Assoc. Petroleum~Geol.} \def \aapgb{\aapg~Bull.} \def \aapgm{\aapg~Memoir} \def \acha{Applied Comput. Harmonic Analysis} \def \actag{Acta~Geophysica} \def \agt{Acta Geologica Taiwanica} \def \amsci{American~Scientist} \def \ajs{Am.~J.~Sci.} \def \angeo{Annals.~Geophy.} \def \areps{Annu.~Rev. Earth Planet.~Sci.} \def \aa{Astron.~Astophys.} \def \ag{Astron.~Geophys.} \def \aj{Astrophys.~J.} \def \araa{Annu.~Rev. Astron.~Astrophys.} \def \bssa{Bull.~Seismol.~Soc.~Am.} \def \basinr{Basin~Research} \def \bvolc{Bull.~Volcanology} \def \beri{Bull.~Earthquake Research~Inst.} \def \biesas{Bull. Inst. Earth Sci., Academia Sinica} \def \bvsj{Bull. Volc. Soc. Japan} \def \ccp{Commun. Comput. Phys.} \def \cse{Computing Science~Engineering} \def \csd{Computational Science \& Discovery} \def \cg{Computers \& Geosciences} \def \comphys{Computers in Physics} \def
  \chemgeo{Chem.~Geol.} \def \cj{Computer Journal} \def \cjes{Can.~J. Earth~Sci.} \def \cnsns{Comm.~Nonlin.~Sci. Num.~Sim.} \def \crst{Cold~Regions Sci.~Tech.} \def \dao{Dynamics~Atmos.~Oceans} \def \ecgeo{Econ.~Geol.} \def \eqspec{Earthquake Spectra} \def \eps{Earth~Planets~Space} \def \esr{Earth-Sci.~Rev.} \def \epsl{Earth~Planet. Sci.~Lett.} \def \eos{Eos~Trans. \agu} \def \fb{First~Break} \def \geol{Geology} \def \gsa{Geol.~Soc.~Am.} \def \gsat{GSA~Today} \def \gsab{Geol.~Soc.~Am. Bull.} \def \geomag{Geol.~Mag.} \def \geop{Geophysics} \def \geos{Geosphere} \def \ggg{Geochem.~Geophy.~Geosyst.} \def \gi{Geof\'isica.~Internacional} \def \geophysj{Geophys.~J.} \def \gji{Geophys.~J.~Int.} \def \grl{Geophys.~Res.~Lett.} \def \gjras{Geophys.~J. R.~Astron.~Soc.} \def \gml{Geo-Marine~Lett.} \def \gms{Geophys.~Monogr.~Series} \def \gp{Geophys.~Prosp.} \def \ia{Island~Arc} \def \ieee{IEEE~Trans.~Vis.~Comp.~Graphics} \def \igr{International Geology Review} \def \ijnme{Int. J. Numerical Meth. Engineering} \def
  \ip{Inverse~Problems} \def \jag{J.~App.~Geophys.} \def \jam{J.~App.~Mech.} \def \jaes{J.~Asian Earth~Sci.} \def \jasa{J.~Acoust.~Soc.~Am.} \def \jastp{J.~Atmos. Solar-Terr.~Phys.} \def \jclim{J.~Climate} \def \jel{J.~Elasticity} \def \jem{J.~Eng.~Mech.} \def \jcp{J.~Comp.~Phys.} \def \jfm{J.~Fluid~Mech.} \def \jgr{J.~Geophys.~Res.} \def \jgrb{J.~Geophys.~Res. Biogeosciences} \def \jgres{J.~Geophys.~Res. Earth~Surface} \def \jgrse{J.~Geophys.~Res. Solid~Earth} \def \jgrsp{J.~Geophys.~Res. Space~Physics} \def \jg{J.~Geodynamics} \def \jgeo{J.~Geology} \def \jgeop{J.~Geophysics} \def \jgsl{J.~Geol.~Soc. London} \def \jseis{J.~Seis.} \def \jmg{J.~Metamorphic~Geol.} \def \jmp{J.~Math.~Phys.} \def \jnaiam{J.~Num.~Analysis, Industrial App. Math.} \def \jota{J.~Optim. Th.~App.} \def \jpe{J.~Phys.~Earth} \def \jrssb{J.~R.~Statist.~Soc.~B} \def \jsaes{J.~South American Earth Sciences} \def \jsc{J.~Sci.~Comput.} \def \jsg{J.~Struct.~Geol.} \def \jsr{J.~Sed.~Res.} \def \jscs{J.~Statist. Comput.~Simul.} \def
  \jvgr{J.~Volcan. Geothermal~Res.} \def \ledge{Leading~Edge} \def \lncs{Lecture~Notes in Computer~Science} \def \lith{Lithosphere} \def \mg{Marine~Geology} \def \mgr{Marine Geophysical Researches} \def \mathgeo{Mathematical Geology} \def \mgsc{Mem. Geol. Soc. China} \def \mi{Math.~Intelligencer} \def \mnras{Mon.~Not. R.~Astron.~Soc.} \def \mpg{Marine Petroleum Geology} \def \mwr{Monthly Weather Review} \def \nat{Nature} \def \natee{Nature Reviews Earth \& Environment} \def \natcom{Nature~Communications} \def \natgeo{Nature~Geoscience} \def \natphys{Nature~Physics} \def \nathaz{Natural~Hazards} \def \nhess{Natural~Hazards and Earth System Sciences} \def \nzjgg{New.~Zealand J.~Geol.~Geophys.} \def \numa{Numerical Algorithms} \def \ps{Polar~Science} \def \pt{Physics~Today} \def \pce{Physics and Chemistry of the Earth} \def \pepi{Phys.~Earth Planet.~Inter.} \def \ptrs{Phil.~Trans. R.~Soc.} \def \ptrsl{\ptrs~Lond.} \def \ptrsA{\ptrs~A.} \def \ptrsla{\ptrs~Lond.~A.} \def \pag{Pure~App.~Geophys.} \def
  \pnas{Proc.~Natl. Acad.~Sci.} \def \prsa{Proc.~R.~Soc.A} \def \pieee{Proc.~IEEE} \def \pers{Photogrammetric Eng. \& Remote Sensing} \def \qjrms{Q.~J.~R.~Meteorol.~Soc.} \def \qsr{Quaternary Sci.~Rev.} \def \rpp{Rep.~Prog.~Phys.} \def \rgsp{Rev.~Geophys.~Space.~Phys.} \def \rgp{Rev.~Geophys.} \def \rmp{Rev.~Mod.~Phys.} \def \sa{Scientific American} \def \se{Solid~Earth} \def \sci{Science} \def \sciadv{Science~Advances} \def \scirep{Scientific Reports} \def \scipro{Science~Progress} \def \srl{Seismol.~Res.~Lett.} \def \spej{Soc.~Petroleum Engineers~J.} \def \sp{Solar~Physics} \def \segea{SEG Expanded Abstracts} \def \seg{Soc. Economic Geologists} \def \sepm{Soc. Sedimentary Geology} \def \sirev{SIAM~Rev.} \def \sjna{SIAM~J. Numer.~Anal.} \def \sjsc{SIAM~J. Sci.~Comp.} \def \sjssc{SIAM~J. Sci.~Stat.~Comp.} \def \statsci{Statistical~Science} \def \survgp{Surv.~Geophys.} \def \tao{Terr.~Atmos. Oceanic~Sci.} \def \tec{Tectonics} \def \ternov{Terra~Nova} \def \tecphy{Tectonophysics} \def \tsr{The~Seismic~Record}
  \def \wm{Wave~Motion} \def \zis{Zisin (J.~Seis. Soc.~Japan)}
\begin{thebibliography}{}

\bibitem[\protect\citeauthoryear{Aki and Richards}{Aki and Richards}{2002}]{aki2002quantitative}
Aki, K. and P.~G. Richards (2002).
\newblock {\em Quantitative seismology}.

\bibitem[\protect\citeauthoryear{Arrowsmith, Trugman, MacCarthy, Bergen, Lumley, and Magnani}{Arrowsmith et~al.}{2022}]{arrowsmith2022big}
Arrowsmith, S.~J., D.~T. Trugman, J.~MacCarthy, K.~J. Bergen, D.~Lumley, and M.~B. Magnani (2022).
\newblock Big data seismology.
\newblock {\em Reviews of Geophysics\/}~{\bf 60\/}(2), e2021RG000769.

\bibitem[\protect\citeauthoryear{Bao, Bielak, Ghattas, Kallivokas, O'Hallaron, Shewchuk, and Xu}{Bao et~al.}{1998}]{bao1998large}
Bao, H., J.~Bielak, O.~Ghattas, L.~F. Kallivokas, D.~R. O'Hallaron, J.~R. Shewchuk, and J.~Xu (1998).
\newblock Large-scale simulation of elastic wave propagation in heterogeneous media on parallel computers.
\newblock {\em Computer methods in applied mechanics and engineering\/}~{\bf 152\/}(1-2), 85--102.

\bibitem[\protect\citeauthoryear{Barker, Chue~Hong, Katz, Lamprecht, Martinez-Ortiz, Psomopoulos, Harrow, Castro, Gruenpeter, Martinez, et~al.}{Barker et~al.}{2022}]{barker2022introducing}
Barker, M., N.~P. Chue~Hong, D.~S. Katz, A.-L. Lamprecht, C.~Martinez-Ortiz, F.~Psomopoulos, J.~Harrow, L.~J. Castro, M.~Gruenpeter, P.~A. Martinez, et~al. (2022).
\newblock Introducing the fair principles for research software.
\newblock {\em Scientific Data\/}~{\bf 9\/}(1), 622.

\bibitem[\protect\citeauthoryear{Beyreuther, Barsch, Krischer, Megies, Behr, and Wassermann}{Beyreuther et~al.}{2010}]{obspy2010}
Beyreuther, M., R.~Barsch, L.~Krischer, T.~Megies, Y.~Behr, and J.~Wassermann (2010).
\newblock {ObsPy: A Python toolbox for seismology}.
\newblock {\em \srl\/}~{\bf 81\/}(3), 530--533.

\bibitem[\protect\citeauthoryear{Bozda{\u{g}}, Trampert, and Tromp}{Bozda{\u{g}} et~al.}{2011}]{bozdaug2011misfit}
Bozda{\u{g}}, E., J.~Trampert, and J.~Tromp (2011).
\newblock Misfit functions for full waveform inversion based on instantaneous phase and envelope measurements.
\newblock {\em Geophysical Journal International\/}~{\bf 185\/}(2), 845--870.

\bibitem[\protect\citeauthoryear{Bozda\u{g}, Peter, Lefebvre, Komatitsch, Tromp, Hill, Podhorszki, and Pugmire}{Bozda\u{g} et~al.}{2016}]{glad15}
Bozda\u{g}, E., D.~Peter, M.~Lefebvre, D.~Komatitsch, J.~Tromp, J.~Hill, N.~Podhorszki, and D.~Pugmire (2016).
\newblock {Global adjoint tomography: first-generation model}.
\newblock {\em \gji\/}~{\bf 207}, 1739--1766.

\bibitem[\protect\citeauthoryear{Brudzinski, Hubenthal, Fasola, and Schnorr}{Brudzinski et~al.}{2021}]{brudzinski2021learning}
Brudzinski, M., M.~Hubenthal, S.~Fasola, and E.~Schnorr (2021).
\newblock Learning in a crisis: Online skill building workshop addresses immediate pandemic needs and offers possibilities for future trainings.
\newblock {\em Seismological Society of America\/}~{\bf 92\/}(5), 3215--3230.

\bibitem[\protect\citeauthoryear{Chow, Kaneko, Tape, Modrak, and Townend}{Chow et~al.}{2020}]{Chow2020}
Chow, B., Y.~Kaneko, C.~Tape, R.~Modrak, and J.~Townend (2020).
\newblock {An automated workflow for adjoint tomography---waveform misfits and synthetic inversions for the North Island, New Zealand}.
\newblock {\em \gji\/}~{\bf 223}, 1461--1480.

\bibitem[\protect\citeauthoryear{Chue~Hong, Katz, Barker, Lamprecht, Martinez, Psomopoulos, Harrow, Castro, Gruenpeter, Martinez, et~al.}{Chue~Hong et~al.}{2022}]{chue2022fair}
Chue~Hong, N.~P., D.~S. Katz, M.~Barker, A.-L. Lamprecht, C.~Martinez, F.~E. Psomopoulos, J.~Harrow, L.~J. Castro, M.~Gruenpeter, P.~A. Martinez, et~al. (2022).
\newblock Fair principles for research software (fair4rs principles).
\newblock {\em Zenodo\/}.

\bibitem[\protect\citeauthoryear{CIG}{CIG}{2016a}]{CIGbestpracticesSOFTWARE}
CIG (2016a).
\newblock {Software Development Best Practices for the CIG Community}.
\newblock {\url{https://github.com/geodynamics/best_practices/blob/master/SoftwareDevelopmentBestPractices.md}} (last accessed July 2023).

\bibitem[\protect\citeauthoryear{CIG}{CIG}{2016b}]{CIGbestpracticesTRAINING}
CIG (2016b).
\newblock {Software Training Best Practices for the CIG Community}.
\newblock {\url{https://github.com/geodynamics/best_practices/blob/master/TrainingBestPractices.md}} (last accessed July 2023).

\bibitem[\protect\citeauthoryear{Denolle, Waldhauser, Tape, Bozdag, and Wang}{Denolle et~al.}{2024}]{Denolle2024}
Denolle, M., F.~Waldhauser, C.~Tape, E.~Bozdag, and I.~Wang (2024, 8).
\newblock Scoped update: a cloud and hpc software platform forcomputational seismology.

\bibitem[\protect\citeauthoryear{Folch, Abril, Afanasiev, Amati, Bader, Badia, Bayraktar, Barsotti, Basili, Bernardi, Boehm, Brizuela, Brogi, Cabrera, Casarotti, Castro, Cerminara, Cirella, Cheptsov, Conejero, Costa, {de la Asunción}, {de la Puente}, Djuric, Dorozhinskii, Espinosa, Esposti-Ongaro, Farnós, Favretto-Cristini, Fichtner, Fournier, Gabriel, Gallard, Gibbons, Glimsdal, González-Vida, Gracia, Gregorio, Gutierrez, Halldorsson, Hamitou, Houzeaux, Jaure, Kessar, Krenz, Krischer, Laforet, Lanucara, Li, Lorenzino, Lorito, Løvholt, Macedonio, Macías, Marín, {Martínez Montesinos}, Mingari, Moguilny, Montellier, Monterrubio-Velasco, Moulard, Nagaso, Nazaria, Niethammer, Pardini, Pienkowska, Pizzimenti, Poiata, Rannabauer, Rojas, Rodriguez, Romano, Rudyy, Ruggiero, Samfass, Sánchez-Linares, Sanchez, Sandri, Scala, Schaeffer, Schuchart, Selva, Sergeant, Stallone, Taroni, Thrastarson, Titos, Tonelllo, Tonini, Ulrich, Vilotte, Vöge, Volpe, {Aniko Wirp}, and Wössner}{Folch et~al.}{2023}]{Folch2023}
Folch, A., C.~Abril, M.~Afanasiev, G.~Amati, M.~Bader, R.~M. Badia, H.~B. Bayraktar, S.~Barsotti, R.~Basili, F.~Bernardi, C.~Boehm, B.~Brizuela, F.~Brogi, E.~Cabrera, E.~Casarotti, M.~J. Castro, M.~Cerminara, A.~Cirella, A.~Cheptsov, J.~Conejero, A.~Costa, M.~{de la Asunción}, J.~{de la Puente}, M.~Djuric, R.~Dorozhinskii, G.~Espinosa, T.~Esposti-Ongaro, J.~Farnós, N.~Favretto-Cristini, A.~Fichtner, A.~Fournier, A.-A. Gabriel, J.-M. Gallard, S.~J. Gibbons, S.~Glimsdal, J.~M. González-Vida, J.~Gracia, R.~Gregorio, N.~Gutierrez, B.~Halldorsson, O.~Hamitou, G.~Houzeaux, S.~Jaure, M.~Kessar, L.~Krenz, L.~Krischer, S.~Laforet, P.~Lanucara, B.~Li, M.~C. Lorenzino, S.~Lorito, F.~Løvholt, G.~Macedonio, J.~Macías, G.~Marín, B.~{Martínez Montesinos}, L.~Mingari, G.~Moguilny, V.~Montellier, M.~Monterrubio-Velasco, G.~E. Moulard, M.~Nagaso, M.~Nazaria, C.~Niethammer, F.~Pardini, M.~Pienkowska, L.~Pizzimenti, N.~Poiata, L.~Rannabauer, O.~Rojas, J.~E. Rodriguez, F.~Romano, O.~Rudyy, V.~Ruggiero, P.~Samfass,
  C.~Sánchez-Linares, S.~Sanchez, L.~Sandri, A.~Scala, N.~Schaeffer, J.~Schuchart, J.~Selva, A.~Sergeant, A.~Stallone, M.~Taroni, S.~Thrastarson, M.~Titos, N.~Tonelllo, R.~Tonini, T.~Ulrich, J.-P. Vilotte, M.~Vöge, M.~Volpe, S.~{Aniko Wirp}, and U.~Wössner (2023).
\newblock The eu center of excellence for exascale in solid earth (cheese): Implementation, results, and roadmap for the second phase.
\newblock {\em Future Generation Computer Systems\/}~{\bf 146}, 47--61.

\bibitem[\protect\citeauthoryear{Gabriel, Ulrich, Marchandon, Biemiller, and Rekoske}{Gabriel et~al.}{2023}]{gabriel2023}
Gabriel, A.-A., T.~Ulrich, M.~Marchandon, J.~Biemiller, and J.~Rekoske (2023).
\newblock 3d dynamic rupture modeling of the 6 february 2023, kahramanmara{\c{s}}, turkey m w 7.8 and 7.7 earthquake doublet using early observations.
\newblock {\em The Seismic Record\/}~{\bf 3\/}(4), 342--356.

\bibitem[\protect\citeauthoryear{Graves, Jordan, Callaghan, Deelman, Field, Juve, Kesselman, Maechling, Mehta, Milner, Okaya, Small, and Vahi}{Graves et~al.}{2011}]{Graves2011}
Graves, R., T.~H. Jordan, S.~Callaghan, E.~Deelman, E.~Field, G.~Juve, C.~Kesselman, P.~Maechling, G.~Mehta, K.~Milner, D.~Okaya, P.~Small, and K.~Vahi (2011).
\newblock Cybershake: A physics-based seismic hazard model for southern california.
\newblock {\em \pag\/}~{\bf 168}, 367--381.

\bibitem[\protect\citeauthoryear{Graves}{Graves}{1998}]{graves1998three}
Graves, R.~W. (1998).
\newblock Three-dimensional finite-difference modeling of the san andreas fault: source parameterization and ground-motion levels.
\newblock {\em Bulletin of the Seismological Society of America\/}~{\bf 88\/}(4), 881--897.

\bibitem[\protect\citeauthoryear{Holtzman, A., J., F., and D.}{Holtzman et~al.}{2018}]{holtzman2018}
Holtzman, B., P.~A., P.~J., W.~F., and R.~D. (2018).
\newblock Machine learning reveals cyclic changes in seismic source spectra in geysers geothermal field.
\newblock {\em Science Advances\/}~{\bf 4\/}(5), eaao2929.

\bibitem[\protect\citeauthoryear{Huppenkothen, Arendt, Hogg, Ram, VanderPlas, and Rokem}{Huppenkothen et~al.}{2018}]{Huppenkothen18}
Huppenkothen, D., A.~Arendt, D.~W. Hogg, K.~Ram, J.~T. VanderPlas, and A.~Rokem (2018).
\newblock Hack weeks as a model for data science education and collaboration.
\newblock {\em Proceedings of the National Academy of Sciences\/}~{\bf 115\/}(36), 8872--8877.

\bibitem[\protect\citeauthoryear{Igel}{Igel}{2017}]{Igel}
Igel, H. (2017).
\newblock {\em {Computational Seismology: A Practical Introduction}}.
\newblock Oxford U. Press.

\bibitem[\protect\citeauthoryear{Jiang and Denolle}{Jiang and Denolle}{2020}]{jiang2020noisepy}
Jiang, C. and M.~A. Denolle (2020).
\newblock Noisepy: A new high-performance python tool for ambient-noise seismology.
\newblock {\em Seismological Research Letters\/}~{\bf 91\/}(3), 1853--1866.

\bibitem[\protect\citeauthoryear{Jones, Okubo, Clements, and Denolle}{Jones et~al.}{2020}]{jones2020seisio}
Jones, J.~P., K.~Okubo, T.~Clements, and M.~A. Denolle (2020).
\newblock Seisio: A fast, efficient geophysical data architecture for the julia language.
\newblock {\em Seismological research letters\/}~{\bf 91\/}(4), 2368--2377.

\bibitem[\protect\citeauthoryear{K{\"a}ser, Castro, Hermann, and Pelties}{K{\"a}ser et~al.}{2010}]{kaser2010seissol}
K{\"a}ser, M., C.~Castro, V.~Hermann, and C.~Pelties (2010).
\newblock Seissol--a software for seismic wave propagation simulations.
\newblock In {\em High Performance Computing in Science and Engineering, Garching/Munich 2009: Transactions of the Fourth Joint HLRB and KONWIHR Review and Results Workshop, Dec. 8-9, 2009, Leibniz Supercomputing Centre, Garching/Munich, Germany}, pp.\  281--292. Springer.

\bibitem[\protect\citeauthoryear{Komatitsch, Liu, Tromp, S\"{u}ss, Stidham, and Shaw}{Komatitsch et~al.}{2004}]{Koma2004}
Komatitsch, D., Q.~Liu, J.~Tromp, P.~S\"{u}ss, C.~Stidham, and J.~H. Shaw (2004).
\newblock {Simulations of ground motion in the Los Angeles basin based upon the spectral-element method}.
\newblock {\em \bssa\/}~{\bf 94\/}(1), 187--206.

\bibitem[\protect\citeauthoryear{Komatitsch, Ritsema, and Tromp}{Komatitsch et~al.}{2002}]{Koma2002}
Komatitsch, D., J.~Ritsema, and J.~Tromp (2002).
\newblock {The spectral-element method, Beowulf computing, and global seismology}.
\newblock {\em \sci\/}~{\bf 298}, 1737--1742.

\bibitem[\protect\citeauthoryear{Komatitsch and Tromp}{Komatitsch and Tromp}{2002a}]{KomaTromp2002a}
Komatitsch, D. and J.~Tromp (2002a).
\newblock {Spectral-element simulations of global seismic wave propagation---I. Validation}.
\newblock {\em \gji\/}~{\bf 149}, 390--412.

\bibitem[\protect\citeauthoryear{Komatitsch and Tromp}{Komatitsch and Tromp}{2002b}]{KomaTromp2002b}
Komatitsch, D. and J.~Tromp (2002b).
\newblock {Spectral-element simulations of global seismic wave propagation---II. Three-dimensional models, oceans, rotation and self-gravitation}.
\newblock {\em \gji\/}~{\bf 150}, 308--318.

\bibitem[\protect\citeauthoryear{Komatitsch and Vilotte}{Komatitsch and Vilotte}{1998}]{KomaVilotte1998}
Komatitsch, D. and J.-P. Vilotte (1998).
\newblock {The spectral element method: An efficient tool to simulate the seismic response of 2D and 3D geological structures}.
\newblock {\em \bssa\/}~{\bf 88\/}(2), 368--392.

\bibitem[\protect\citeauthoryear{Kong, Trugman, Ross, Bianco, Meade, and Gerstoft}{Kong et~al.}{2019}]{kong2019machine}
Kong, Q., D.~T. Trugman, Z.~E. Ross, M.~J. Bianco, B.~J. Meade, and P.~Gerstoft (2019).
\newblock Machine learning in seismology: Turning data into insights.
\newblock {\em Seismological Research Letters\/}~{\bf 90\/}(1), 3--14.

\bibitem[\protect\citeauthoryear{Krauss, Ni, Henderson, and Denolle}{Krauss et~al.}{2023}]{krauss2023seismology}
Krauss, Z., Y.~Ni, S.~Henderson, and M.~Denolle (2023).
\newblock Seismology in the cloud: guidance for the individual researcher.
\newblock {\em Seismica\/}~{\bf 2\/}(2).

\bibitem[\protect\citeauthoryear{Krenz, Uphoff, Ulrich, Gabriel, Abrahams, Dunham, and Bader}{Krenz et~al.}{2021}]{krenz_SC2021}
Krenz, L., C.~Uphoff, T.~Ulrich, A.-A. Gabriel, L.~S. Abrahams, E.~M. Dunham, and M.~Bader (2021).
\newblock 3d acoustic-elastic coupling with gravity: the dynamics of the 2018 palu, sulawesi earthquake and tsunami.
\newblock In {\em Proceedings of the International Conference for High Performance Computing, Networking, Storage and Analysis}, SC '21, New York, NY, USA. Association for Computing Machinery.

\bibitem[\protect\citeauthoryear{Krischer, Aiman, Batholomaus, Donner, {van Driel}, Duru, Garina, Gessele, Gunawan, Hable, Hadziioannou, Koymans, Leeman, Lindner, Ling, Mengies, Nunn, Rijal, Salvermoser, Soza, Tape, Taufiqurrahman, Vargas, Wassermann, W\"olfl, Williams, Wollherr, and Igel}{Krischer et~al.}{2018}]{seismolive}
Krischer, L., Y.~A. Aiman, T.~Batholomaus, S.~Donner, M.~{van Driel}, K.~Duru, K.~Garina, K.~Gessele, T.~Gunawan, S.~Hable, C.~Hadziioannou, M.~Koymans, J.~Leeman, F.~Lindner, A.~Ling, T.~Mengies, C.~Nunn, A.~Rijal, J.~Salvermoser, S.~T. Soza, C.~Tape, T.~Taufiqurrahman, D.~Vargas, J.~Wassermann, F.~W\"olfl, M.~Williams, S.~Wollherr, and H.~Igel (2018).
\newblock {{\em seismo-live}: An educational online library of Jupyter notebooks for seismology}.
\newblock {\em \srl\/}~{\bf 89\/}(6), 2413--2419.

\bibitem[\protect\citeauthoryear{Liu and Gu}{Liu and Gu}{2012}]{LiuGu2012}
Liu, Q. and Y.~J. Gu (2012).
\newblock {Seismic imaging: From classical to adjoint tomography}.
\newblock {\em \tecphy\/}~{\bf 566-567}, 31--66.

\bibitem[\protect\citeauthoryear{MacCarthy, Marcillo, and Trabant}{MacCarthy et~al.}{2020}]{maccarthy2020seismology}
MacCarthy, J., O.~Marcillo, and C.~Trabant (2020).
\newblock Seismology in the cloud: A new streaming workflow.
\newblock {\em Seismological Research Letters\/}~{\bf 91\/}(3), 1804--1812.

\bibitem[\protect\citeauthoryear{Marru, Gunathilake, Herath, Tangchaisin, Pierce, Mattmann, Singh, Gunarathne, Chinthaka, Gardler, Slominski, Douma, Perera, and Weerawarana}{Marru et~al.}{2011}]{airavata}
Marru, S., L.~Gunathilake, C.~Herath, P.~Tangchaisin, M.~Pierce, C.~Mattmann, R.~Singh, T.~Gunarathne, E.~Chinthaka, R.~Gardler, A.~Slominski, A.~Douma, S.~Perera, and S.~Weerawarana (2011).
\newblock Apache airavata: a framework for distributed applications and computational workflows.
\newblock In {\em Proceedings of the 2011 ACM Workshop on Gateway Computing Environments}, GCE '11, New York, NY, USA, pp.\  21–28. Association for Computing Machinery.

\bibitem[\protect\citeauthoryear{McLennan and Kennell}{McLennan and Kennell}{2010}]{HUBzero}
McLennan, M. and R.~Kennell (2010).
\newblock Hubzero: A platform for dissemination and collaboration in computational science and engineering.
\newblock {\em Computing in Science \& Engineering\/}~{\bf 12\/}(2), 48--53.

\bibitem[\protect\citeauthoryear{Modrak, Borisov, Lefebvre, and Tromp}{Modrak et~al.}{2018}]{seisflows}
Modrak, R.~T., D.~Borisov, M.~Lefebvre, and J.~Tromp (2018).
\newblock {SeisFlows---Flexible waveform inversion software}.
\newblock {\em \cg\/}~{\bf 115}, 88--95.

\bibitem[\protect\citeauthoryear{Morra, Bozda\u{g}, Knepley, Räss, and Vesselinov}{Morra et~al.}{2021}]{Morra-Eos2021}
Morra, G., E.~Bozda\u{g}, M.~Knepley, L.~Räss, and V.~Vesselinov (2021).
\newblock A tectonic shift in analytics and computing is coming.
\newblock {\em Eos\/}, 102.

\bibitem[\protect\citeauthoryear{Mousavi and Beroza}{Mousavi and Beroza}{2022}]{mousavi2022deep}
Mousavi, S.~M. and G.~C. Beroza (2022).
\newblock Deep-learning seismology.
\newblock {\em Science\/}~{\bf 377\/}(6607), eabm4470.

\bibitem[\protect\citeauthoryear{Nakata, Gualtieri, and Fichtner}{Nakata et~al.}{2019}]{nakata2019seismic}
Nakata, N., L.~Gualtieri, and A.~Fichtner (2019).
\newblock {\em Seismic ambient noise}.
\newblock Cambridge University Press.

\bibitem[\protect\citeauthoryear{{NCEDC}}{{NCEDC}}{2014}]{NCEDC}
{NCEDC} (2014).
\newblock Northern california earthquake data center.
\newblock Dataset.

\bibitem[\protect\citeauthoryear{Ni, Denolle, Fatland, Alterman, Lipovsky, and Knuth}{Ni et~al.}{2023}]{Ni23}
Ni, Y., M.~A. Denolle, R.~Fatland, N.~Alterman, B.~P. Lipovsky, and F.~Knuth (2023, 10).
\newblock {An Object Storage for Distributed Acoustic Sensing}.
\newblock {\em Seismological Research Letters\/}~{\bf 95\/}(1), 499--511.

\bibitem[\protect\citeauthoryear{Olsen and Archuleta}{Olsen and Archuleta}{1996}]{olsen1996three}
Olsen, K.~B. and R.~J. Archuleta (1996).
\newblock Three-dimensional simulation of earthquakes on the los angeles fault system.
\newblock {\em Bulletin of the Seismological Society of America\/}~{\bf 86\/}(3), 575--596.

\bibitem[\protect\citeauthoryear{Pelties, de~la Puente, Ampuero, Brietzke, and Käser}{Pelties et~al.}{2012}]{pelties_threedimensional_2012}
Pelties, C., J.~de~la Puente, J.-P. Ampuero, G.~B. Brietzke, and M.~Käser (2012).
\newblock Three-dimensional dynamic rupture simulation with a high-order discontinuous galerkin method on unstructured tetrahedral meshes.
\newblock {\em Journal of Geophysical Research: Solid Earth\/}~{\bf 117\/}(B2).

\bibitem[\protect\citeauthoryear{Pelties, Gabriel, and Ampuero}{Pelties et~al.}{2014}]{pelties_verification_2014}
Pelties, C., A.-A. Gabriel, and J.-P. Ampuero (2014).
\newblock Verification of an ader-dg method for complex dynamic rupture problems.
\newblock {\em Geoscientific Model Development\/}~{\bf 7\/}(3), 847--866.

\bibitem[\protect\citeauthoryear{P{\'e}rez and Granger}{P{\'e}rez and Granger}{2007}]{perez2007ipython}
P{\'e}rez, F. and B.~E. Granger (2007).
\newblock Ipython: a system for interactive scientific computing.
\newblock {\em Computing in science \& engineering\/}~{\bf 9\/}(3), 21--29.

\bibitem[\protect\citeauthoryear{Peter, Komatitsch, Luo, Martin, {Le Goff}, Casarotti, {Le Loher}, Magnoni, Liu, Blitz, Nissen-Meyer, Basini, and Tromp}{Peter et~al.}{2011}]{DPeter2011}
Peter, D., D.~Komatitsch, Y.~Luo, R.~Martin, N.~{Le Goff}, E.~Casarotti, P.~{Le Loher}, F.~Magnoni, Q.~Liu, C.~Blitz, T.~Nissen-Meyer, P.~Basini, and J.~Tromp (2011).
\newblock {Forward and adjoint simulations of seismic wave propagation on fully unstructured hexahedral meshes}.
\newblock {\em \gji\/}~{\bf 186}, 721--739.

\bibitem[\protect\citeauthoryear{Pimentel, Murta, Braganholo, and Freire}{Pimentel et~al.}{2019}]{Pimentel19}
Pimentel, J.~F., L.~Murta, V.~Braganholo, and J.~Freire (2019).
\newblock A large-scale study about quality and reproducibility of jupyter notebooks.
\newblock In {\em 2019 IEEE/ACM 16th International Conference on Mining Software Repositories (MSR)}, pp.\  507--517.

\bibitem[\protect\citeauthoryear{Plesch, Shaw, Benson, Bryant, Carena, Cooke, Dolan, Fuis, Gath, Grant, Hauksson, Jordan, Kamerling, Legg, Lindvall, Magistrale, Nicholson, Niemi, Oskin, Perry, Planansky, Rockwell, Shearer, Sorlien, Süss, Suppe, Treiman, and Yeats}{Plesch et~al.}{2007}]{Plesch2007}
Plesch, A., J.~H. Shaw, C.~Benson, W.~A. Bryant, S.~Carena, M.~Cooke, J.~Dolan, G.~Fuis, E.~Gath, L.~Grant, E.~Hauksson, T.~Jordan, M.~Kamerling, M.~Legg, S.~Lindvall, H.~Magistrale, C.~Nicholson, N.~Niemi, M.~Oskin, S.~Perry, G.~Planansky, T.~Rockwell, P.~Shearer, C.~Sorlien, M.~P. Süss, J.~Suppe, J.~Treiman, and R.~Yeats (2007, 12).
\newblock {Community Fault Model (CFM) for Southern California}.
\newblock {\em Bulletin of the Seismological Society of America\/}~{\bf 97\/}(6), 1793--1802.

\bibitem[\protect\citeauthoryear{Rost and Thomas}{Rost and Thomas}{2009}]{rost2009improving}
Rost, S. and C.~Thomas (2009).
\newblock Improving seismic resolution through array processing techniques.
\newblock {\em Surveys in geophysics\/}~{\bf 30}, 271--299.

\bibitem[\protect\citeauthoryear{Savran, Bayona, Iturrieta, Asim, Bao, Bayliss, Herrmann, Schorlemmer, Maechling, and Werner}{Savran et~al.}{2022}]{savran2022pycsep}
Savran, W.~H., J.~A. Bayona, P.~Iturrieta, K.~M. Asim, H.~Bao, K.~Bayliss, M.~Herrmann, D.~Schorlemmer, P.~J. Maechling, and M.~J. Werner (2022).
\newblock pycsep: a python toolkit for earthquake forecast developers.
\newblock {\em Seismological Society of America\/}~{\bf 93\/}(5), 2858--2870.

\bibitem[\protect\citeauthoryear{Sawi, Holtzman, Walter, and Paisley}{Sawi et~al.}{2022}]{sawi2022}
Sawi, T., B.~Holtzman, F.~Walter, and J.~Paisley (2022).
\newblock An unsupervised machine-learning approach to understanding seismicity at an alpine glacier.
\newblock {\em Journal of Geophysical Research: Earth Surface\/}~{\bf 127\/}(12), e2022JF006909.

\bibitem[\protect\citeauthoryear{SeisSCOPED}{SeisSCOPED}{2024}]{HPS2024}
SeisSCOPED, W.~I. (2024).
\newblock {\em High-Performance Seismology}.
\newblock SeisSCOPED.
\newblock Dynamic textbook.

\bibitem[\protect\citeauthoryear{Shearer}{Shearer}{2019}]{shearer2019introduction}
Shearer, P.~M. (2019).
\newblock {\em Introduction to seismology}.
\newblock Cambridge university press.

\bibitem[\protect\citeauthoryear{Small, Gill, Maechling, Taborda, Callaghan, Jordan, Olsen, Ely, and Goulet}{Small et~al.}{2017}]{Small2017}
Small, P., D.~Gill, P.~J. Maechling, R.~Taborda, S.~Callaghan, T.~H. Jordan, K.~B. Olsen, G.~P. Ely, and C.~Goulet (2017, 09).
\newblock {The SCEC Unified Community Velocity Model Software Framework}.
\newblock {\em Seismological Research Letters\/}~{\bf 88\/}(6), 1539--1552.

\bibitem[\protect\citeauthoryear{Stanzione, West, Evans, Minyard, Ghattas, and Panda}{Stanzione et~al.}{2020}]{Frontera2020}
Stanzione, D., J.~West, R.~T. Evans, T.~Minyard, O.~Ghattas, and D.~K. Panda (2020).
\newblock Frontera: The evolution of leadership computing at the national science foundation.
\newblock In {\em Practice and Experience in Advanced Research Computing 2020: Catch the Wave}, PEARC '20, New York, NY, USA, pp.\  106–111. Association for Computing Machinery.

\bibitem[\protect\citeauthoryear{Stein and Wysession}{Stein and Wysession}{2009}]{stein2009introduction}
Stein, S. and M.~Wysession (2009).
\newblock {\em An introduction to seismology, earthquakes, and earth structure}.
\newblock John Wiley \& Sons.

\bibitem[\protect\citeauthoryear{Stubbs, Cardone, Packard, Jamthe, Padhy, Terry, Looney, Meiring, Black, Dahan, Cleveland, and Jacobs}{Stubbs et~al.}{2021}]{Tapis}
Stubbs, J., R.~Cardone, M.~Packard, A.~Jamthe, S.~Padhy, S.~Terry, J.~Looney, J.~Meiring, S.~Black, M.~Dahan, S.~Cleveland, and G.~Jacobs (2021).
\newblock Tapis: An api platform for reproducible, distributed computational research.
\newblock In K.~Arai (Ed.), {\em Advances in Information and Communication}, Cham, pp.\  878--900. Springer International Publishing.

\bibitem[\protect\citeauthoryear{Tape, Bozdag, Denolle, Waldhauser, and Wang}{Tape et~al.}{2022}]{scoped2022NSFposter}
Tape, C., E.~Bozdag, M.~Denolle, F.~Waldhauser, and I.~Wang (2022).
\newblock {SCOPED: Seismic COmputational Platform for Empowering Discovery [Year 1]}.
\newblock {Zenodo. \url{https://doi.org/10.5281/zenodo.6862979}}.

\bibitem[\protect\citeauthoryear{Tape, Liu, Maggi, and Tromp}{Tape et~al.}{2009}]{Tape2009}
Tape, C., Q.~Liu, A.~Maggi, and J.~Tromp (2009).
\newblock {Adjoint tomography of the southern California crust}.
\newblock {\em Science\/}~{\bf 325}, 988--992.

\bibitem[\protect\citeauthoryear{Thurin, Braunmiller, Cardozo, Ding, Liu, McPherson, Modrak, and Tape}{Thurin et~al.}{2023}]{Thurin2023SSAabstract}
Thurin, J., J.~Braunmiller, F.~R.~R. Cardozo, L.~Ding, Q.~Liu, A.~McPherson, R.~Modrak, and C.~Tape (2023).
\newblock {MTUQ: A high-performance Python package for moment tensor estimation and uncertainty quantification}.
\newblock {Abstract at 2023 SSA Annual Meeting, San Juan, Puerto Rico, 17-20 April}.

\bibitem[\protect\citeauthoryear{Tromp}{Tromp}{2020}]{Tromp2020}
Tromp, J. (2020).
\newblock Seismic wavefield imaging of earth’s interior across scales.
\newblock {\em Nat Rev Earth Environ\/}~{\bf 1}, 40--53.

\bibitem[\protect\citeauthoryear{Tromp, Tape, and Liu}{Tromp et~al.}{2005}]{Tromp2005}
Tromp, J., C.~Tape, and Q.~Liu (2005).
\newblock {Seismic tomography, adjoint methods, time reversal, and banana-doughnut kernels}.
\newblock {\em \gji\/}~{\bf 160}, 195--216.

\bibitem[\protect\citeauthoryear{Uphoff, Krenz, Ulrich, Wolf, Knoll, David, Li, Dorozhinskii, Heinecke, Wollherr, Bohn, Schliwa, Brietzke, Taufiqurrahman, Anger, Rettenberger, Simonis, Gabriel, Pauw, Breuer, Kutschera, Hendrawan~Palgunadi, Rannabauer, van~de Wiel, Li, Chamberlain, Yun, Rekoske, G, and Bader}{Uphoff et~al.}{2024}]{uphoff_seissol_2024}
Uphoff, C., L.~Krenz, T.~Ulrich, S.~Wolf, A.~Knoll, S.~David, D.~Li, R.~Dorozhinskii, A.~Heinecke, S.~Wollherr, M.~Bohn, N.~Schliwa, G.~Brietzke, T.~Taufiqurrahman, S.~Anger, S.~Rettenberger, F.~Simonis, A.~Gabriel, V.~Pauw, A.~Breuer, F.~Kutschera, K.~Hendrawan~Palgunadi, L.~Rannabauer, L.~van~de Wiel, B.~Li, C.~Chamberlain, J.~Yun, J.~Rekoske, Y.~G, and M.~Bader (2024, May).
\newblock Seissol.

\bibitem[\protect\citeauthoryear{Waldhauser and Ellsworth}{Waldhauser and Ellsworth}{2000}]{WaldhauserEllsworth2000}
Waldhauser, F. and W.~L. Ellsworth (2000).
\newblock {A double-difference earthquake location algorithm: Method and application to the Northern Hayward fault, California}.
\newblock {\em \bssa\/}~{\bf 90\/}(6), 1353--1368.

\bibitem[\protect\citeauthoryear{Waldhauser and Schaff}{Waldhauser and Schaff}{2008}]{waldhauser2008large}
Waldhauser, F. and D.~P. Schaff (2008).
\newblock Large-scale relocation of two decades of northern california seismicity using cross-correlation and double-difference methods.
\newblock {\em Journal of Geophysical Research: Solid Earth\/}~{\bf 113\/}(B8).

\bibitem[\protect\citeauthoryear{Wang, Bozdag, denolle, and Waldhauser}{Wang et~al.}{2023}]{Wang2023}
Wang, I., E.~Bozdag, denolle, and F.~Waldhauser (2023, 9).
\newblock Scoped: Seismic computational platform for empowering discovery [year 2].

\bibitem[\protect\citeauthoryear{Wang, Waldhauser, Schaff, Tolstoy, Wilcock, and Tan}{Wang et~al.}{2024}]{Wang2024}
Wang, K., F.~Waldhauser, D.~Schaff, M.~Tolstoy, W.~S.~D. Wilcock, and Y.~J. Tan (2024, 07).
\newblock Real‐time detection of volcanic unrest and eruption at axial seamount using machine learning.
\newblock {\em Seismological Research Letters\/}~{\bf 95\/}(5), 2651--2662.

\bibitem[\protect\citeauthoryear{Wang, Evans, and Huang}{Wang et~al.}{2019}]{Wang_container}
Wang, Y., R.~T. Evans, and L.~Huang (2019).
\newblock Performant container support for hpc applications.
\newblock In {\em Practice and Experience in Advanced Research Computing 2019: Rise of the Machines (Learning)}, PEARC '19, New York, NY, USA. Association for Computing Machinery.

\bibitem[\protect\citeauthoryear{Wang, Pavlis, Yang, and Ma}{Wang et~al.}{2022}]{wang_mspass_2022}
Wang, Y., G.~L. Pavlis, W.~Yang, and J.~Ma (2022).
\newblock {MsPASS}: A data management and processing framework for seismology.
\newblock {\em Seismological Research Letters\/}~{\bf 93\/}(1), 426--434.

\bibitem[\protect\citeauthoryear{Yu, Bhaskaran, Chen, Ross, Hauksson, and Clayton}{Yu et~al.}{2021}]{yu21scedc}
Yu, E., A.~Bhaskaran, S.~Chen, Z.~E. Ross, E.~Hauksson, and R.~W. Clayton (2021, 06).
\newblock Southern california earthquake data now available in the aws cloud.
\newblock {\em Seismological Research Letters\/}~{\bf 92\/}(5), 3238--3247.

\bibitem[\protect\citeauthoryear{Yuan, Ni, Lin, and Denolle}{Yuan et~al.}{2023}]{yuan2023better}
Yuan, C., Y.~Ni, Y.~Lin, and M.~Denolle (2023).
\newblock Better together: Ensemble learning for earthquake detection and phase picking.
\newblock {\em IEEE Transactions on Geoscience and Remote Sensing\/}.

\bibitem[\protect\citeauthoryear{Zhu and Rivera}{Zhu and Rivera}{2002}]{ZhuRivera2002}
Zhu, L. and L.~A. Rivera (2002).
\newblock {A note on the dynamic and static displacements from a point source in multilayered media}.
\newblock {\em \gji\/}~{\bf 148}, 619--627.

\bibitem[\protect\citeauthoryear{Zhu, Hou, Yang, Datta, Mousavi, Ellsworth, and Beroza}{Zhu et~al.}{2023}]{zhu2023quakeflow}
Zhu, W., A.~B. Hou, R.~Yang, A.~Datta, S.~M. Mousavi, W.~L. Ellsworth, and G.~C. Beroza (2023).
\newblock Quakeflow: a scalable machine-learning-based earthquake monitoring workflow with cloud computing.
\newblock {\em Geophysical Journal International\/}~{\bf 232\/}(1), 684--693.

\end{thebibliography}

\def \agu{Am.~Geophys.~Un.} \def \usgs{U.S.~Geol.~Survey} \def \dggs{Alaska Div. Geol. Geophys. Surv.} \def \antsci{Antarctic~Science} \def \aapg{Am.~Assoc. Petroleum~Geol.} \def \aapgb{\aapg~Bull.} \def \aapgm{\aapg~Memoir} \def \acha{Applied Comput. Harmonic Analysis} \def \actag{Acta~Geophysica} \def \agt{Acta Geologica Taiwanica} \def \amsci{American~Scientist} \def \ajs{Am.~J.~Sci.} \def \angeo{Annals.~Geophy.} \def \areps{Annu.~Rev. Earth Planet.~Sci.} \def \aa{Astron.~Astophys.} \def \ag{Astron.~Geophys.} \def \aj{Astrophys.~J.} \def \araa{Annu.~Rev. Astron.~Astrophys.} \def \bssa{Bull.~Seismol.~Soc.~Am.} \def \basinr{Basin~Research} \def \bvolc{Bull.~Volcanology} \def \beri{Bull.~Earthquake Research~Inst.} \def \biesas{Bull. Inst. Earth Sci., Academia Sinica} \def \bvsj{Bull. Volc. Soc. Japan} \def \ccp{Commun. Comput. Phys.} \def \cse{Computing Science~Engineering} \def \csd{Computational Science \& Discovery} \def \cg{Computers \& Geosciences} \def \comphys{Computers in Physics} \def
  \chemgeo{Chem.~Geol.} \def \cj{Computer Journal} \def \cjes{Can.~J. Earth~Sci.} \def \cnsns{Comm.~Nonlin.~Sci. Num.~Sim.} \def \crst{Cold~Regions Sci.~Tech.} \def \dao{Dynamics~Atmos.~Oceans} \def \ecgeo{Econ.~Geol.} \def \eqspec{Earthquake Spectra} \def \eps{Earth~Planets~Space} \def \esr{Earth-Sci.~Rev.} \def \epsl{Earth~Planet. Sci.~Lett.} \def \eos{Eos~Trans. \agu} \def \fb{First~Break} \def \geol{Geology} \def \gsa{Geol.~Soc.~Am.} \def \gsat{GSA~Today} \def \gsab{Geol.~Soc.~Am. Bull.} \def \geomag{Geol.~Mag.} \def \geop{Geophysics} \def \geos{Geosphere} \def \ggg{Geochem.~Geophy.~Geosyst.} \def \gi{Geof\'isica.~Internacional} \def \geophysj{Geophys.~J.} \def \gji{Geophys.~J.~Int.} \def \grl{Geophys.~Res.~Lett.} \def \gjras{Geophys.~J. R.~Astron.~Soc.} \def \gml{Geo-Marine~Lett.} \def \gms{Geophys.~Monogr.~Series} \def \gp{Geophys.~Prosp.} \def \ia{Island~Arc} \def \ieee{IEEE~Trans.~Vis.~Comp.~Graphics} \def \igr{International Geology Review} \def \ijnme{Int. J. Numerical Meth. Engineering} \def
  \ip{Inverse~Problems} \def \jag{J.~App.~Geophys.} \def \jam{J.~App.~Mech.} \def \jaes{J.~Asian Earth~Sci.} \def \jasa{J.~Acoust.~Soc.~Am.} \def \jastp{J.~Atmos. Solar-Terr.~Phys.} \def \jclim{J.~Climate} \def \jel{J.~Elasticity} \def \jem{J.~Eng.~Mech.} \def \jcp{J.~Comp.~Phys.} \def \jfm{J.~Fluid~Mech.} \def \jgr{J.~Geophys.~Res.} \def \jgrb{J.~Geophys.~Res. Biogeosciences} \def \jgres{J.~Geophys.~Res. Earth~Surface} \def \jgrse{J.~Geophys.~Res. Solid~Earth} \def \jgrsp{J.~Geophys.~Res. Space~Physics} \def \jg{J.~Geodynamics} \def \jgeo{J.~Geology} \def \jgeop{J.~Geophysics} \def \jgsl{J.~Geol.~Soc. London} \def \jseis{J.~Seis.} \def \jmg{J.~Metamorphic~Geol.} \def \jmp{J.~Math.~Phys.} \def \jnaiam{J.~Num.~Analysis, Industrial App. Math.} \def \jota{J.~Optim. Th.~App.} \def \jpe{J.~Phys.~Earth} \def \jrssb{J.~R.~Statist.~Soc.~B} \def \jsaes{J.~South American Earth Sciences} \def \jsc{J.~Sci.~Comput.} \def \jsg{J.~Struct.~Geol.} \def \jsr{J.~Sed.~Res.} \def \jscs{J.~Statist. Comput.~Simul.} \def
  \jvgr{J.~Volcan. Geothermal~Res.} \def \ledge{Leading~Edge} \def \lncs{Lecture~Notes in Computer~Science} \def \lith{Lithosphere} \def \mg{Marine~Geology} \def \mgr{Marine Geophysical Researches} \def \mathgeo{Mathematical Geology} \def \mgsc{Mem. Geol. Soc. China} \def \mi{Math.~Intelligencer} \def \mnras{Mon.~Not. R.~Astron.~Soc.} \def \mpg{Marine Petroleum Geology} \def \mwr{Monthly Weather Review} \def \nat{Nature} \def \natee{Nature Reviews Earth \& Environment} \def \natcom{Nature~Communications} \def \natgeo{Nature~Geoscience} \def \natphys{Nature~Physics} \def \nathaz{Natural~Hazards} \def \nhess{Natural~Hazards and Earth System Sciences} \def \nzjgg{New.~Zealand J.~Geol.~Geophys.} \def \numa{Numerical Algorithms} \def \ps{Polar~Science} \def \pt{Physics~Today} \def \pce{Physics and Chemistry of the Earth} \def \pepi{Phys.~Earth Planet.~Inter.} \def \ptrs{Phil.~Trans. R.~Soc.} \def \ptrsl{\ptrs~Lond.} \def \ptrsA{\ptrs~A.} \def \ptrsla{\ptrs~Lond.~A.} \def \pag{Pure~App.~Geophys.} \def
  \pnas{Proc.~Natl. Acad.~Sci.} \def \prsa{Proc.~R.~Soc.A} \def \pieee{Proc.~IEEE} \def \pers{Photogrammetric Eng. \& Remote Sensing} \def \qjrms{Q.~J.~R.~Meteorol.~Soc.} \def \qsr{Quaternary Sci.~Rev.} \def \rpp{Rep.~Prog.~Phys.} \def \rgsp{Rev.~Geophys.~Space.~Phys.} \def \rgp{Rev.~Geophys.} \def \rmp{Rev.~Mod.~Phys.} \def \sa{Scientific American} \def \se{Solid~Earth} \def \sci{Science} \def \sciadv{Science~Advances} \def \scirep{Scientific Reports} \def \scipro{Science~Progress} \def \srl{Seismol.~Res.~Lett.} \def \spej{Soc.~Petroleum Engineers~J.} \def \sp{Solar~Physics} \def \segea{SEG Expanded Abstracts} \def \seg{Soc. Economic Geologists} \def \sepm{Soc. Sedimentary Geology} \def \sirev{SIAM~Rev.} \def \sjna{SIAM~J. Numer.~Anal.} \def \sjsc{SIAM~J. Sci.~Comp.} \def \sjssc{SIAM~J. Sci.~Stat.~Comp.} \def \statsci{Statistical~Science} \def \survgp{Surv.~Geophys.} \def \tao{Terr.~Atmos. Oceanic~Sci.} \def \tec{Tectonics} \def \ternov{Terra~Nova} \def \tecphy{Tectonophysics} \def \tsr{The~Seismic~Record}
  \def \wm{Wave~Motion} \def \zis{Zisin (J.~Seis. Soc.~Japan)}

\end{document}